\documentclass[10pt,twocolumn,letterpaper]{article}

\usepackage[pagenumbers]{cvpr} % To force page numbers, e.g. for an arXiv version

\usepackage{self}

\definecolor{cvprblue}{rgb}{0.21,0.49,0.74}
\usepackage[pagebackref,breaklinks,colorlinks,allcolors=cvprblue]{hyperref}

\def\paperID{*****} % *** Enter the Paper ID here
\def\confName{CVPR}
\def\confYear{2026}

\title{Is This Edit Correct? A Multi-Dimensional Benchmark\\ for Reasoning-Aware Image Editing}

\author{
    \textbf{Yixuan Ding$^{\spadesuit}$} \quad
    \textbf{Wei Huang$^{\diamondsuit}$} \quad
    \textbf{Ruijie Quan$^{\spadesuit}$}\textsuperscript{,†} \quad
    \textbf{Xiaojuan Qi$^{\diamondsuit}$} \quad
    \textbf{Yi Yang$^{\spadesuit}$} \\
    $^{\spadesuit}$ Zhejiang University \quad
    $^{\diamondsuit}$ The University of Hong Kong \\
    \
    \faGithub\hspace{0.2em}
    \texttt{GitHub: }
    \href{https://github.com/Yixuan-Ding-ZJU/RE-Edit}{\texttt{github.com/Yixuan-Ding-ZJU/RE-Edit}} \\
    \raisebox{-0.25em}{\includegraphics[height=1.2em]{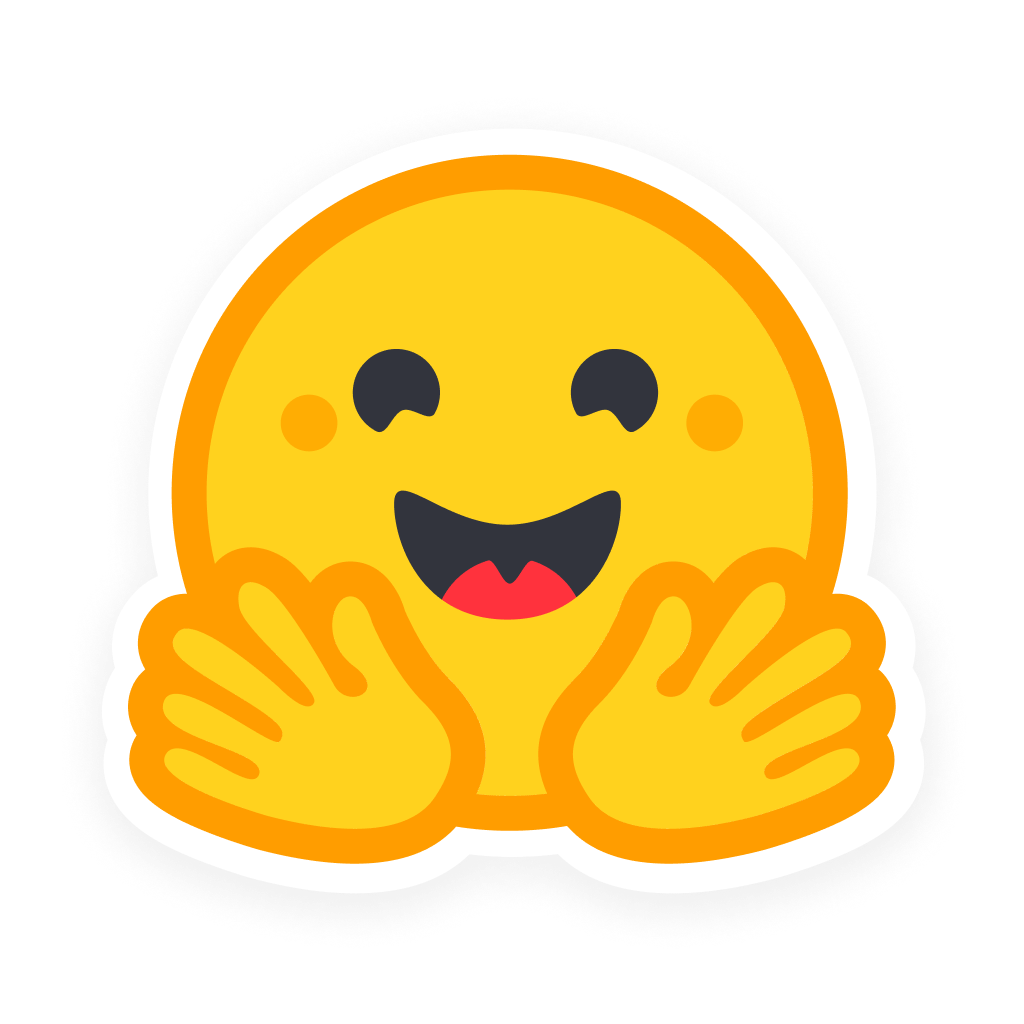}}
    \texttt{Benchmark: }
    \href{https://huggingface.co/datasets/Yixuan-Ding-ZJU/RE-Edit}{\texttt{huggingface.co/datasets/Yixuan-Ding-ZJU/RE-Edit}}
}

\begin{document}

\twocolumn[
    
    \maketitle
    \vspace{-15pt}
    \noindent\parbox{\textwidth}{%
    \centering
    \includegraphics[width=\textwidth]{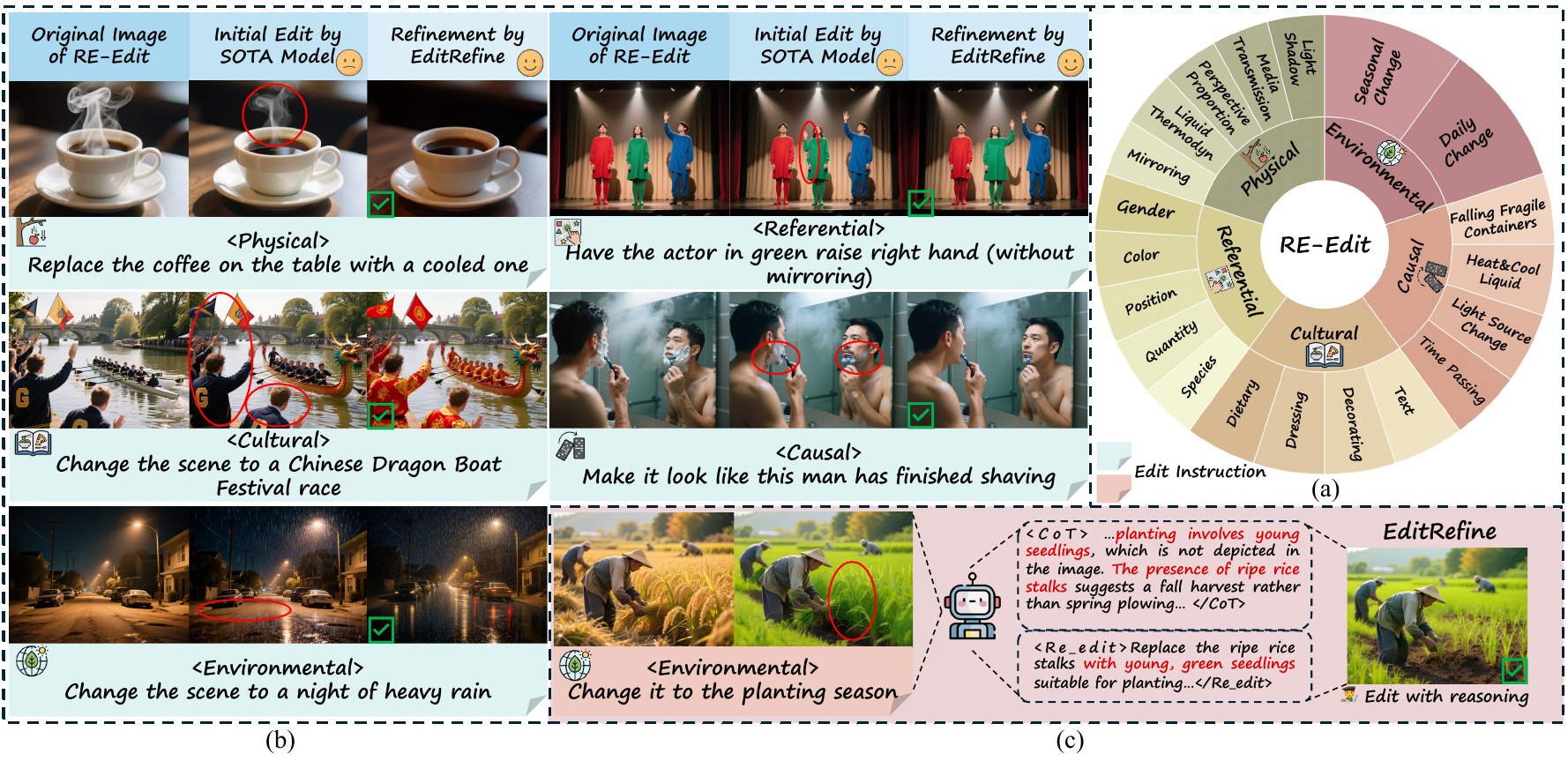}\par
    \captionsetup{skip=1pt} % 这里越小，caption 越贴近图片
    \captionof{figure}{\textbf{RE-Edit benchmark and EditRefine overview.} (a) Human-logic–derived taxonomy across five reasoning dimensions. (b) Representative RE-Edit cases: SOTA failures (red) and EditRefine corrections (green). (c) EditRefine pipeline: diagnose and generate refined re-edit instruction for execution.}
    \label{fig:representative case}%
    \vspace{10pt}
  }
  ]

\begin{abstract}

Diffusion-based image editing has achieved strong visual fidelity under natural language instructions, yet most existing systems still operate at the level of surface instruction following, without reasoning about the implicit contextual constraints embedded in real user requests. This often leads to visually plausible but logically inconsistent edits.
In this work, we introduce \textbf{RE-Edit}, a benchmark for \textbf{RE}asoning-aware image \textbf{Edit}ing that evaluates image editing systems across five complementary reasoning dimensions: \textit{\textbf{physical}}, \textit{\textbf{environmental}}, \textit{\textbf{cultural}}, \textit{\textbf{causal}}, and \textit{\textbf{referential}}. RE-Edit comprises 1,000 carefully curated samples, each designed such that visual plausibility alone is insufficient and correct editing requires satisfying implicit logical constraints. To support fine-grained analysis, we establish dimension-aligned evaluation criteria and conduct a comprehensive study of ten open-source and two commercial image editing models. Our results show that even advanced systems frequently struggle with implicit multi-dimensional reasoning despite producing high-quality visuals. We further present a lightweight reasoning-guided post-edit baseline as an initial exploration, illustrating how inserting explicit reasoning can help mitigate such failures in a model-agnostic manner.
\end{abstract}    
\vspace{-15pt}
\section{Introduction}
% --- Logic Step 1: The Contrast (LLM Reasoning vs. Image Editing) & The Gap ---
Recent advances in Large Language Models (LLMs)~\cite{jaech2024openai-o1,guo2025deepseekr1} and Multimodal LLMs (MLLMs)~\cite{bai2025qwen3vltechnicalreport-qwen3vl-aigc-mllm,hurst2024gpt-4o,chen2025scaling} have substantially improved machine reasoning abilities. 
Supported by paradigms like Chain-of-Thought (CoT), these models can decompose complex tasks, infer latent intentions, and perform multi-step logical reasoning. In contrast, while generative image editing~\cite{zhu2017toward-IE,isola2017image-IE,wu2025qwen-image,rombach2022highresolutionimagesynthesislatent-IE-LDM,zhang2025context,song2025insert,brooks2023instructpix2pixlearningfollowimage-IE-InstructPix2Pix, xu2024gg, shen2025tarpro} has achieved impressive progress in visual fidelity and controllability driven by diffusion-based frameworks~\cite{diffusion_algorithm, song2020score-based-generative}, it still predominantly operates at the level of surface instruction following.

Specifically, this surface-level instruction following arises because most existing image editing systems are optimized to align textual cues with visual appearances, directly translating instructions into pixel-level transformations. This paradigm is effective when instructions specify \emph{explicit visual attributes}, \eg, ``\texttt{render a man without shaving foam and beard}'' (Figure~\ref{fig:representative case}).
However, many real user instructions are inherently \emph{implicit}, conveying desired edits through logic reasoning rather than literal descriptions, \eg, ``\texttt{make it look like this man has finished shaving}''.
Executing such an edit correctly requires inferring a latent thought chain, \ie, shaving process finished$\rightarrow$removal of shaving foam$\rightarrow$removal facial stubble, rather than manipulating isolated visual elements alone.
As a result, visually plausible yet logically inconsistent edits frequently emerge, exposing a critical gap between \textbf{instruction following} and \textbf{reasoning-aware image editing}.
%without explicitly modeling the contextual constraints implied by an edit.

% Specifically, current image editing systems typically operate on a correlation-driven basis, \textcolor{blue}{mapping textual patterns directly to visual appearances.} This approach suffices when instructions provide \textit{explicit visual descriptions}, e.g.,\texttt{``render an oxidized, yellow apple''}). However, a critical limitation emerges when instructions rely on \textit{implicit reasoning}—a scenario where the user implies a visual change through causal or contextual logic rather than literal description, e.g.,\texttt{``make the apple look like it has been left out for a while''}. In such cases, correctly executing the edit requires inferring a latent causal chain prolonged air exposure $\rightarrow$ oxidation $\rightarrow$ discoloration/shriveling. \textcolor{blue}{Existing models, prioritizing statistical texture matching over logical consistency, frequently fail to bridge this gap, revealing a significant disconnect between \textbf{\textit{instruction following}} and \textbf{\textit{reasoning-aware}} capabilities.}

% --- Logic Step 2: Human Cognitive Workflow & RE-Edit Benchmark ---

To better analyze its underlying causes, we take inspiration from the editing workflow of human professionals.
When interpreting abstract instructions, human editors typically consider multiple aspects of an edit, \ie, \ding{172} identifying the intended target (\textit{referential}), \ding{173} preserving physical plausibility (\textit{physical}), \ding{174} adapting changes to the surrounding scene (\textit{environmental}), \ding{175} respecting social and cultural conventions (\textit{cultural}), and \ding{176} maintaining causal or event-level coherence (\textit{causal}).
To this end, we raise a question: \textbf{\textit{to what extent do current image editing systems account for these implicit considerations?}} However, most existing benchmarks~\cite{wang2023imageneditoreditbenchadvancing-editbench,sheynin2023emueditpreciseimage-emuedit,ye2025imgeditunifiedimageediting-imgeditbench, liu2025step1xeditpracticalframeworkgeneral-aigc-geditbench-IE-step-edit-v1p1, ye2025unicedit10mdatasetbenchmarkbreaking-unicbench, wu2025krisbenchbenchmarkingnextlevelintelligent} primarily focus on visual fidelity or instruction compliance, and do not explicitly evaluate the kinds of implicit, multi-dimensional human logic reasoning considerations described above.
% This motivates the need for a dedicated benchmark that can systematically evaluate reasoning-aware image editing.

To answer this question, we introduce \textbf{RE-Edit}, a benchmark designed to systematically evaluate \textbf{RE}asoning-aware image \textbf{Edit}ing across multiple implicit reasoning dimensions. 
RE-Edit comprises 1,000 carefully curated samples spanning five fundamental reasoning dimensions, \ie, \textit{\textbf{physical}}, \textit{\textbf{environmental}}, \textit{\textbf{cultural}}, \textit{\textbf{causal}}, and \textit{\textbf{referential}}. Each sample is explicitly designed to probe a specific reasoning requirement, such that visual plausibility alone is insufficient and correct editing depends on satisfying the associated implicit logical constraints.
Since such reasoning failures may not be reflected by conventional perceptual or task-level metrics~\cite{ku2024viescoreexplainablemetricsconditional-viescore}, we further establish \textbf{dimension-aligned evaluation criteria} to support fine-grained assessment on RE-Edit, and conduct a comprehensive evaluation of state-of-the-art image editing systems, including open-source models, \ie, Janus-4o~\cite{chen2025sharegpt4oimagealigningmultimodalmodels-janus-7B}, FLUX.1.Kontext~\cite{labs2025flux1kontextflowmatching-IE-kontext}, Step1X-Edit-v1p1\&v1p2-preview~\cite{liu2025step1xeditpracticalframeworkgeneral-aigc-geditbench-IE-step-edit-v1p1, yin2025reasoneditreasoningenhancedimageediting-step-edit-v1p2}, DreamOmni2~\cite{xia2025dreamomni2multimodalinstructionbasedediting-dreamomni2}, Ovis-U1-3B~\cite{wang2025ovisu1technicalreport-ovis-u1-3b}, HiDream-E1~\cite{cai2025hidreami1highefficientimagegenerative-hidream-e1},Qwen-Image-Edit~\cite{wu2025qwen-image}, FLUX.2 Dev and commercial models, \ie, Nano Banana~\cite{comanici2025gemini}, and Seedream 4.0~\cite{seedream2025seedream40nextgenerationmultimodal-seedream4}. 

%Results suggest that several state-of-the-art editors achieve high visual quality yet still tend to reasoning mismatches under implicit constraints, and that post-edit refinement can improve reasoning correctness in a plug-and-play manner.

Beyond evaluation, we further explore whether adding explicit reasoning signals can help mitigate the reasoning failures revealed by RE-Edit.
As an initial exploration, we implement a simple reasoning-guided baseline, termed \textbf{EditRefine}, which operates as a post-edit refinement step on top of existing image editing models.
EditRefine leverages an MLLM-based reasoning agent, optimized via reinforcement learning, to diagnose potential logical inconsistencies in the generated edits and synthesize refined editing instructions, while keeping the underlying generative model unchanged.
Rather than serving as a comprehensive solution, EditRefine is intended as a proof-of-concept baseline that illustrates how explicit reasoning can be incorporated in a lightweight and model-agnostic manner.
Our contributions are threefold:

\begin{itemize}[topsep=1pt,itemsep=2pt,parsep=0pt,leftmargin=2em]
    \item We formalize \textbf{reasoning-aware image editing} as a distinct capability beyond surface-level instruction following, and characterize it through a multi-dimensional taxonomy inspired by human editing workflows.
    \item We introduce \textbf{RE-Edit}, a benchmark of 1,000 curated samples spanning five complementary reasoning dimensions, together with dimension-aligned evaluation criteria for systematically assessing implicit reasoning in image editing.
    %\item We introduce \textbf{RE-Edit}, a benchmark of 1,000 curated samples spanning five complementary reasoning dimensions, enabling systematic evaluation of implicit reasoning in image editing with dimension-aligned evaluation criteria.
    \item We conduct a comprehensive and systematic evaluation of 12 state-of-the-art image editing systems on RE-Edit, providing detailed analyses of reasoning performance across dimensions, and include a lightweight reasoning-guided post-edit framework.
    %\item We implement \textbf{EditRefine}, a lightweight reasoning-aware post-edit framework that performs instruction-level diagnosis and second-pass refinement to mitigate logic-dependent failures without fine-tuning the underlying editor backbone.
\end{itemize}

%%%%%%%%%%%%%%%%%%%%%%%%%%%%%%%%%%%%%%%%%%%%%%%%%%%%%%
% Related Works Section 
%\vspace{-4pt} % Add by Yixuan
\section{Related Works}
\subsection{Image Editing Models}
\label{subsec:imgedit_models}
Diffusion models now dominate high-fidelity image editing, reshaping the field from latent-space manipulation to natural-language-driven editing. Early diffusion-based methods primarily rely on {inversion-based techniques}, which map an input image into the diffusion latent space to preserve structure during editing. Representative approaches include noise-perturbation methods such as SDEdit~\cite{meng2022sdeditguidedimagesynthesis-sdedit} and trajectory-based DDIM inversion~\cite{song2022denoisingdiffusionimplicitmodels-DDIM}, as well as optimization-based variants like Null-text inversion~\cite{mokady2022nulltextinversioneditingreal-nulltext}, and PTI~\cite{dong2023prompttuninginversiontextdriven-pti}. To support localized or structured modifications, these methods are often combined with {attention-based mechanisms}, including Prompt-to-Prompt~\cite{hertz2022prompttopromptimageeditingcross-p2p}, Plug-and-Play~\cite{tumanyan2022plugandplaydiffusionfeaturestextdriven-pnp}, and Pix2Pix-Zero~\cite{parmar2023zeroshotimagetoimagetranslation-pix2pixzero}.
More recent work emphasizes \textbf{instruction-driven image editing}, training models to directly follow natural-language edit commands. InstructPix2Pix~\cite{brooks2023instructpix2pixlearningfollowimage-IE-InstructPix2Pix} is an early representative, followed by stronger open-source systems such as FLUX.1.Kontext~\cite{labs2025flux1kontextflowmatching-IE-kontext} and Qwen-Image-Edit~\cite{wu2025qwen-image}, as well as commercial models including Nano Banana~\cite{comanici2025gemini} and Seedream 4.0~\cite{seedream2025seedream40nextgenerationmultimodal-seedream4}. While these approaches significantly improve visual fidelity and instruction compliance, they primarily focus on learning statistical alignments between text and appearance instead of  explicit implementation of multi-dimensional reasoning consistency during editing, an aspect systematically studied in our RE-Edit benchmark.

\subsection{Image Editing Benchmarks}
\label{subsec:editbench}
Evaluating image editing is challenging due to subjective visual quality and the difficulty of measuring instruction satisfaction. Early benchmarks such as EditBench~\cite{wang2023imageneditoreditbenchadvancing-editbench} focus on inpainting and simple attribute edits, while later benchmarks, including EmuEdit~\cite{sheynin2023emueditpreciseimage-emuedit} and AnyEdit~\cite{yu2025anyeditmasteringunifiedhighquality-anyedit}, broaden task coverage and adopt automatic evaluation based on pixel similarity or CLIP~\cite{radford2021learningtransferablevisualmodels-clip} scores. More recent efforts, such as ImgEditBench~\cite{ye2025imgeditunifiedimageediting-imgeditbench} and GEdit-Bench~\cite{liu2025step1xeditpracticalframeworkgeneral-aigc-geditbench-IE-step-edit-v1p1}, leverage VLM-based evaluation to assess higher-level semantic compliance.
A few recent benchmarks begin to explore reasoning-related aspects of image editing. 
UnicBench~\cite{ye2025unicedit10mdatasetbenchmarkbreaking-unicbench} associates reasoning with \emph{complex editing}, emphasizing structurally involved operations such as multi-object manipulation and viewpoint changes, while KRIS-Bench~\cite{wu2025krisbenchbenchmarkingnextlevelintelligent} focuses on explicit knowledge-based reasoning via an educational taxonomy (\eg, Bloom’s taxonomy). In contrast, \textbf{RE-Edit} focuses on \emph{implicit, editor-centric logical constraints} that arise in everyday editing requests, and systematically evaluates reasoning behavior across multiple dimensions grounded in human editing workflows. As such, RE-Edit complements existing benchmarks by providing a unified and fine-grained view of reasoning-aware image editing.

% Evaluating image editing is difficult due to subjective visual quality and the challenge of measuring instruction satisfaction. Early benchmarks such as EditBench~\cite{wang2023imageneditoreditbenchadvancing-editbench} focus on inpainting and simple attribute edits. EmuEdit~\cite{sheynin2023emueditpreciseimage-emuedit} and AnyEdit~\cite{yu2025anyeditmasteringunifiedhighquality-anyedit} broaden coverage, but largely rely on CLIP-based~\cite{radford2021learningtransferablevisualmodels-clip} metrics or pixel-wise distances that may miss logic-related failures.

% Recent benchmarks adopt VLM-based evaluation for semantic alignment, including ImgEditBench~\cite{ye2025imgeditunifiedimageediting-imgeditbench} and GEdit-Bench~\cite{liu2025step1xeditpracticalframeworkgeneral-aigc-geditbench-IE-step-edit-v1p1}. UnicBench~\cite{ye2025unicedit10mdatasetbenchmarkbreaking-unicbench} further touches on reasoning, though complex logic remains under-covered.

% KRIS-Bench~\cite{wu2025krisbenchbenchmarkingnextlevelintelligent} targets knowledge-based reasoning using an educational taxonomy (\eg, Bloom's), categorizing tasks into explicit academic disciplines. By contrast, \textbf{RE-Edit} centers on \textit{implicit logical constraints} in editing requests and uses a human-logic taxonomy grounded in the cognitive checks editors perform, complementing subject-oriented benchmarks.

%%%%%%%%%%%%%%%%%%%%%%%%%%%%%%%%%%%%%%%%%%%%%%%%%%%%%%
%%%%%%%%%%%%%%%%%%%%%%%%%%%%%%%%%%%%%%%%%%%%%%%%%%%%%%

%%%%%%%%%%%%%%%%%%%%%%%%%%%%%%%%%%%%%%%%%%%%%%%%%%%%%%
% Figure
\begin{figure*}[!t]
  \centering
  \includegraphics[width=\textwidth]{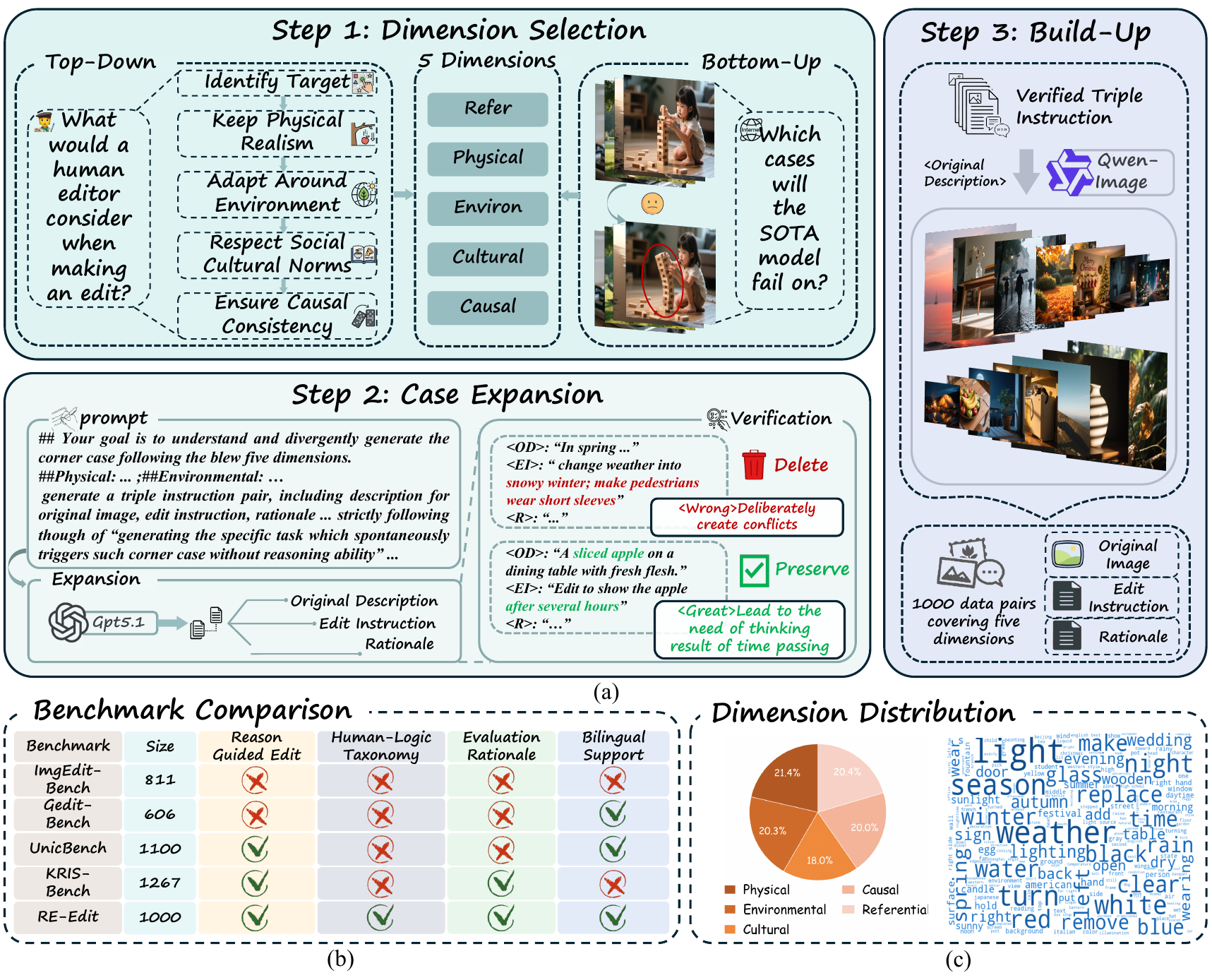}
    \captionsetup{skip=1pt} % 这里越小，caption 越贴近图片
    \captionof{figure}{\textbf{RE-Edit benchmark construction and statistics.} (a) Curation pipeline: define five human-logic reasoning dimensions, expand corner cases, and verify instruction triples. (b) Benchmark comparison on reasoning-guided edits, human-logic taxonomy, evaluation rationales, and bilingual support. (c) RE-Edit dimension distribution and edit-instruction word cloud (generic terms removed).}
  \label{fig:data curation pipeline & benchmark}
  \vspace{-15pt}
\end{figure*}
%%%%%%%%%%%%%%%%%%%%%%%%%%%%%%%%%%%%%%%%%%%%%%%%%%%%%%

% RE-Edit Section
\section{RE-Edit Benchmark}
To evaluate editing scenarios that require implicit reasoning beyond surface instruction alignment, we construct \textbf{RE-Edit}, a benchmark of 1,000 curated samples. 
RE-Edit is organized around a five-dimension taxonomy inspired by how human editors interpret edit requests, and each case is paired with rationale that support fine-grained evaluation.

\subsection{Reasoning Dimensions}
RE-Edit adopts a five-dimensional taxonomy that captures common sources of reasoning failures in instruction-based image editing. 
These dimensions are designed to reflect whether an edited result remains consistent with implicit constraints of the visual world, beyond surface instruction alignment.
Concretely, we formalize such constraints into five complementary reasoning dimensions (Figure~\ref{fig:data curation pipeline & benchmark}).

% RE-Edit adopts a taxonomy derived from how human editors interpret an edit request: they implicitly identify the target entity and ensure the edited result remains consistent with latent constraints of the visual world. Concretely, we formalize these constraints into five complementary dimensions that capture common sources of reasoning failures in instruction-based image editing (Figure~\ref{fig:data curation pipeline & benchmark}).

\noindent\textbf{Physical Consistency.}
Edits should respect fundamental physical constraints of real-world scenes, such as geometric structure, lighting and shading coherence, and other material-dependent effects. Violations of physical consistency often appear visually plausible at a glance, yet contradict basic physical cues upon closer inspection.

\noindent\textbf{Environmental Consistency.}Edits should remain compatible with the broader surrounding context, requiring local modifications (\eg, clothing, accessories, objects) to align with global environmental conditions such as weather, season, or time of day.
The key challenge lies in maintaining scene-level coherence rather than performing isolated foreground or background changes.

% Edits should align with the surrounding context, such that local elements (\eg, clothes, umbrella should be compatible with global environmental conditions change (\eg, scene-wide temporal or weather context). The key challenge is maintaining coherent context rather than isolated foreground or background modification.

\noindent\textbf{Cultural Consistency.}
Edits should respect culturally grounded semantics, where objects, symbols, language, and visual styles are appropriate for the implied cultural or social setting.
Failures often arise when models produce edits that appear visually reasonable in isolation, but violate culturally specific norms implied by the scene context.

% Edits should respect culturally grounded semantics, where objects, symbols, languages, and styles are appropriate for the implied cultural setting. Failures often arise when models default to generic patterns that conflict with context-specific conventions.

\noindent\textbf{Causal Consistency.}
Edits should reflect plausible cause--effect relations implied by the instruction. 
Rather than performing a literal transformation, the model is expected to infer and render the visual consequences of an underlying action, process, or event described implicitly in the request.

% The model must render the consequence of an action or process, rather than producing a literal but logically incomplete transformation.

\noindent\textbf{Referential Consistency.}
Edits should be applied to the intended target in complex scenes, requiring accurate grounding and disambiguation based on language and visual attributes. Common failures include modifying the wrong instance or unintentionally affecting non-target entities.

These dimensions provide a compact and complementary framework for analyzing reasoning-related errors in instruction-based image editing models.

\subsection{Data Curation}
Given the taxonomy above, we instantiate RE-Edit via a human-in-the-loop data curation pipeline that emphasizes logical validity and reasoning depth:

%\begin{enumerate}[label=(\roman*)]
\noindent\textbf{(i) Case Expansion:}$_{\!\!\!}$
We first design a set of dimension-aligned$_{\!}$ seed$_{\!\!}$ cases$_{\!\!}$ that$_{\!\!}$ target$_{\!\!\!}$ representative reasoning challenges under each category. Starting from these seeds, we prompt GPT-5.1 to expand them into (\texttt{original description, edit instruction, rationale}) triples that require implicit inference beyond literal attribute insertion. All generated cases are then manually reviewed to remove ill-posed samples (\eg, deliberate contradictions, ambiguous targets, or incorrect rationales), retaining only coherent edit requests with valid rationales that capture the intended implicit constraint.

% From dimension-aligned corner-case seeds, we prompt GPT-5.1 to expand each seed into \texttt{(original description, edit instruction, rationale)} triples that require implicit inference beyond literal attribute insertion. We then manually filter out ill-posed or trivial samples (\eg, deliberate conflicts, ambiguous targets, or incorrect rationales), retaining only coherent edit requests with valid rationales that capture the key implicit constraint.

\noindent\textbf{(ii) Build Up:} 
For validated triples, we synthesize high-quality images from the \texttt{original description} using Qwen-Image~\cite{wu2025qwen-image}. All images are generated at a fixed resolution of $1472\times1104$. This strategy enables controllable image generation, and facilitates coverage of rare or complex editing scenarios that are difficult to obtain from existing datasets.

The resulting benchmark contains 1,000 samples, each consisting of \texttt{original image}, reasoning-intensive \texttt{edit instruction}, \texttt{rationale}, and annotations (dimension, difficulty), balanced across the five dimensions. More detailed examples are provided in Figure~\ref{fig:reedit_samples}.

\subsection{Evaluation Metrics}
We evaluate instruction-based image editing models from two perspectives: \emph{reasoning correctness} and \emph{general editing quality} (non-reasoning). For general editing quality, we adopt Semantic Consistency (SC) from VIEScore~\cite{ku2024viescoreexplainablemetricsconditional-viescore} and Instruction Following (IF) from UnicEdit~\cite{ye2025unicedit10mdatasetbenchmarkbreaking-unicbench}. SC measures preservation of semantic content, while IF evaluates how well the output literally satisfies the explicit editing instruction; both are scored on a 0-10 scale. Following VIEScore, we define the final SC score as the minimum across its sub-tasks to enforce a strict preservation criterion.
To assess reasoning correctness, we introduce dimension-aligned evaluation criteria tailored to the five-dimensional reasoning of RE-Edit. For each case, the provided \texttt{rationale} highlights the key logical requirement and anchors evaluation to the corresponding constraint. We employ a binary Pass/Fail judgment per case to reflect the discrete nature of logical validity, and report category scores by passing rate. All evaluation prompts and protocols are provided in Appendix~\ref{subsec:evaluation prompt}.

% Reasoning-Guided EditRefine Section
\section{Reasoning-Guided Post-Edit (EditRefine)}
\subsection{Framework}
As an initial exploration, we implement a plug-and-play post-edit refinement framework, termed \textbf{EditRefine}, to examine whether explicit reasoning signals can help mitigate reasoning failures revealed by RE-Edit.
EditRefine operates as a second-pass refinement module: given an initial edited result from a base editor, it performs diagnostic reasoning and applies a corrective re-edit when necessary.

EditRefine consists of two conceptual components: 
(i) a \textbf{Reasoning Agent}, instantiated by a vision-language model capable of multi-step reasoning, and 
(ii) an \textbf{Execution Engine}, implemented using an off-the-shelf diffusion-based image editor.
In our implementation, we use Qwen2.5-VL-7B~\cite{bai2025qwen25vltechnicalreport-aigc-qwen25vl} as the Reasoning Agent and Qwen-Image-Edit~\cite{wu2025qwen-image} as the Execution Engine.
Given the original image, the initial edited result, and the user instruction, the Reasoning Agent performs a CoT-style diagnostic analysis to identify potential logical inconsistencies in the edit (prompt details are provided in Appendix~\ref{subsec:prompt for editrefine}), and then produces a \textit{refined instruction} that explicitly encodes the missing or violated constraints.
The Execution Engine subsequently applies this refined instruction to generate the final output. The inference cost and latency analysis are provided in Appendix \ref{app:inference_cost_analysis}.
Since EditRefine operates purely at the instruction level, it does not require retraining or modification of the diffusion backbone, making it readily applicable to a wide range of image editing systems.

% We propose a plug-and-play framework that adds reasoning to existing instruction-based image editing pipelines by coupling an MLLM with a diffusion editor. The framework serves as a post-edit refinement module: given an initially edited result from a base editor, it performs a second-pass correction guided by diagnostic reasoning.

% EditRefine consists of two components: a \textbf{Reasoning Agent} (a fine-tuned  Qwen2.5-VL-7B~\cite{bai2025qwen25vltechnicalreport-aigc-qwen25vl}) and an \textbf{Execution Engine} (an off-the-shelf Qwen-Image-Edit model~\cite{wu2025qwen-image}). Given the original image, the initial edit, and the user instruction, the Reasoning Agent generates a CoT-based diagnosis that identifies potential logical inconsistencies and produces a Refined Instruction. Details of the prompting format are provided in Appendix~\ref{subsec:prompt for editrefine}. The Execution Engine then applies the Refined Instruction to produce the final output. Because EditRefine operates at the instruction level, it can correct logic-dependent failures without retraining or modifying the diffusion backbone, making it broadly applicable across existing editing pipelines.

\subsection{Training}

To obtain a reasonably capable Reasoning Agent, we adopt a simple two-stage training strategy commonly used in recent LLM work~\cite{guo2025deepseekr1}, consisting of supervised fine-tuning followed by reinforcement learning.
%We emphasize that this training setup is not optimized for performance, but is sufficient to demonstrate the potential benefit of explicit reasoning in post-edit refinement.

% We train the Reasoning Agent with a two-stage pipeline, following common practice in LLMs~\cite{guo2025deepseekr1}: (i) Supervised Fine-Tuning to align the output format and basic diagnostic behavior, and (ii) Reinforcement Learning to improve constraint-sensitive reasoning.

\noindent\textbf{Supervised Fine-Tuning.}
Starting from Qwen2.5-VL-7B, we perform parameter-efficient supervised fine-tuning on instruction-edit pairs augmented with reasoning traces.
This stage aligns the agent with the desired output protocol, enabling it to produce structured \texttt{<CoT>} diagnostics and corresponding \texttt{<Re\_edit>} instructions, and serves as a stable initialization for subsequent reinforcement learning.

% Starting from Qwen2.5-VL-7B, we perform parameter-efficient fine-tuning on instruction--edit pairs augmented with reasoning traces. This stage teaches the agent to produce structured \texttt{<CoT>} diagnostics and \texttt{<Re\_edit>} refined instructions in the required protocol, providing a stable initialization for RL.

\noindent\textbf{Reinforcement Learning.}
Building on the supervised initialization, we employ a dimension-aware reward that prioritizes the most salient reasoning error in each case, preventing reward dilution across multiple dimensions.
Specifically, we use a Max-Deviation strategy to focus training on the dominant reasoning failure, combined with a simple format reward to ensure structured outputs.
%We further improve the agent using Group Relative Policy Optimization~\cite{shao2024deepseekmathpushinglimitsmathematical-grpo}.
%To avoid reward dilution across multiple reasoning dimensions, we adopt a simple max-deviation heuristic that focuses training on the most salient reasoning error in each case.
% During training, refined instructions are executed by the image editor and evaluated with dimension-specific scores, and the reward combines the largest deviation from baseline with a lightweight format constraint.
Additional training details are provided in Appendix~\ref{sec:appendix_training_details}.

% We further optimize the agent using Group Relative Policy Optimization~\cite{shao2024deepseekmathpushinglimitsmathematical-grpo}. Since reasoning correctness is assessed across multiple dimensions, we introduce a Max-Deviation Strategy to target the most critical dimension of reasoning correctness for each training case. By identifying the dimension $d \in D$ with the largest magnitude deviation between its score $S_d$ and baseline $B_d$, we isolate the dominant signal. This prevents reward dilution associated with aggregating multiple, potentially irrelevant dimension scores. Concretely, during online training we execute the refined instruction via Qwen-Image-Edit executor and obtain dimension-specific scores from an evaluator. The reward is defined as:
% {%
%   \setlength{\abovedisplayskip}{6pt}
%   \setlength{\belowdisplayskip}{6pt}
%   \setlength{\abovedisplayshortskip}{6pt}
%   \setlength{\belowdisplayshortskip}{6pt}
%   \[
%     R = \max_{d \in D} \lvert S_d - B_d \rvert + \lambda_{fmt}\cdot R_{fmt}
%   \]
% }
% where $S_d$ and $B_d$ denote the score and baseline for dimension $d$, respectively, and $R_{fmt}$ is a rule-based format reward that enforces structural adherence of the agent. Additional implementation details are provided in Appendix~\ref{sec:appendix_training_details}.

%%%%%%%%%%%%%%%%%%%%%%%%%%%%%%%%%%%%%%%%%%%%%%%%%%%%%%
% Figure
\begin{figure*}[!p]
  \centering
  \includegraphics[width=\textwidth]{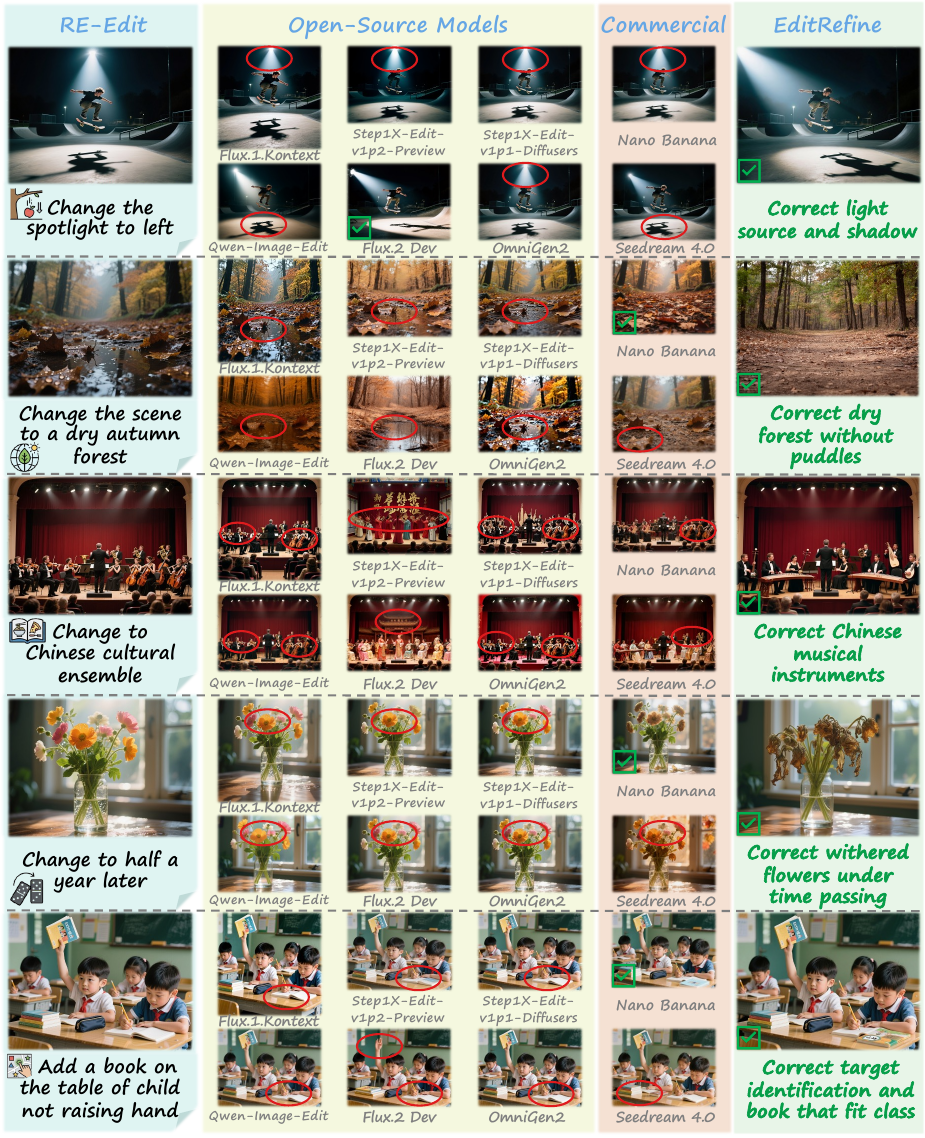}
  \caption{\textbf{Qualitative comparisons on RE-Edit across five reasoning dimensions with EditRefine.} We show representative RE-Edit cases evaluated with strong open-source and commercial image editors, where SOTA outputs often violate human-logic constraints (red circles). \textbf{EditRefine} performs reasoning-guided refinement and produces corrected results (green check marks) by refining initial edits generated by a frozen Qwen-Image-Edit backbone.}
  \label{fig:qualitative case}
\end{figure*}

%\FloatBarrier
\begin{table*}[!t]
\captionsetup{skip=4pt}
\caption{\textbf{Main results on RE-Edit.} Representative open-source and commercial editors evaluated on five reasoning dimensions and two general metrics (IF, SC) by Qwen3-VL-30B; \texttt{Executor-F} and \texttt{Executor-Q} denote the FLUX.2 Dev and Qwen-Image-Edit executors, respectively. Red \textcolor{red!70}{$\uparrow$}
 indicates absolute improvement over the corresponding backbone.}
  \label{tab:main_results}
  \centering
  \resizebox{\textwidth}{!}{%
  %\tiny
  % 字体设计
  \scriptsize
  \setlength\tabcolsep{8pt}
  \renewcommand{\arraystretch}{0.8}
  \begin{tabular}{
    l
    | c c c c c
    | c c 
  }
    \toprule
    \multirow{2}{*}{\textbf{Model}} &
    \multicolumn{5}{c|}{\textbf{Reasoning}} &
    \multicolumn{2}{c}{\textbf{Non-Reasoning}} \\
    \cmidrule(lr){2-6}
    \cmidrule(lr){7-8}
    &
    Physical & Environmental & Cultural & Causal & Referential &
    IF & SC   \\
    \midrule

    \multicolumn{8}{c}{\textit{\color{gray}{Evaluator: Qwen3-VL-30B}}} \\
    \midrule
    \rowcolor{blue!10}\multicolumn{8}{c}{\textit{{Open-source Models}}} \\

Janus-4o &
6.0 & 0.98 & 0.0 & 0.5 & 7.32 &
3.6 & 4.0  \\

FLUX.1.Kontext &
15.4 & 4.4 & 2.8 & 7.5 & 35.3 &
5.8 & 3.9  \\

Step1X-Edit-v1p1 &
15.8 & 7.4 & 2.2 & 7.5 & 38.2 &
6.6 & 3.9  \\

Step1X-Edit-v1p2-preview &
16.4 & 9.4 & 0.6 & 5.0 & 47.5 &
7.0 & 3.9  \\

DreamOmni2 &
15.4 & 9.4 & 2.2 & 9.0 & 36.8 &
6.0 & 3.5 \\

OmniGen2 &
14.5 & 3.4 & 1.1 & 5.5 & 38.2 &
5.4 & 3.6  \\

Ovis-U1-3B &
16.8 & 13.3 & 2.2 & 4.5 & 39.2 &
7.2 & 4.5  \\

HiDream-E1 &
14.0 & 8.9 & 1.1 & 5.5 & 32.4 &
6.0 & 4.8  \\

Qwen-Image-Edit &
21.0 & 12.8 & 3.3 & 13.0 & 50.0 &
7.0 & 3.8 \\

FLUX.2 Dev &
20.1 & 14.8 & 7.8 & 15.0 & 50.5 &
7.7 & 4.0  \\

    \rowcolor{cyan!10}
    \multicolumn{8}{c}{\textit{{Commercial Models}}} \\
Nano Banana &
20.1 & 10.8 & 3.9 & 15.5 & 48.0 &
7.6 & 3.6  \\
Seedream 4.0 &
18.7 & 16.3 & 6.1 & 14.5 & 53.4 &
8.0 & 3.7  \\

    \rowcolor{red!8}
    \multicolumn{8}{c}{\textit{{Plug-In EditRefine}}} \\
    
    Qwen-Image-Edit &
     21.0 & 12.8 & 3.3 & 13.0 & 50.0 &
     7.0 & 3.8    \\
    \quad \textbf{+ EditRefine w Executor-F} &
    21.0\gain{-} & 
    13.8\gain{1.0} &
    6.1\gain{2.8} &
    17.5\gain{4.5} &
    45.6 &
    7.5\gain{0.5} &
    4.2\gain{0.4} \\
    \midrule
        
    FLUX.2 Dev &
    20.1 & 14.8 & 7.8 & 15.0 & 50.5 &
    7.7 & 4.0    \\
    \quad \textbf{+ EditRefine w Executor-Q} &
        22.0\gain{1.9} &
        14.8\gain{-} &
        8.3\gain{0.5} &
        16.0\gain{1.0} &
        48.5 &
        7.7\gain{-} &
        4.4\gain{0.4}  \\
    \midrule

    FLUX.1.Kontext &
    15.4 & 4.4 & 2.8 & 7.5 & 35.3 &
    5.8 & 3.9    \\
    \quad \textbf{+ EditRefine w Executor-Q} &
    14.0 &
    6.9\gain{2.5} &
    4.4\gain{1.6} &
    10.0\gain{2.5}&
    36.8\gain{1.5} &
        6.6\gain{0.8} &
        4.3\gain{0.4}  \\
    \midrule
    Nano Banana &
20.1 & 10.8 & 3.9 & 15.5 & 48.0 &
7.6 & 3.6  \\
    \quad \textbf{+ EditRefine w Executor-Q} &
    19.6 &
    11.8\gain{1.0} &
    6.1\gain{2.2} &
    16.5\gain{1.0}&
    49.5 \gain{1.5}&
        7.8\gain{0.2} &
        3.8\gain{0.2}  \\
    \midrule
    
    Seedream 4.0 &
18.7 & 16.3 & 6.1 & 14.5 & 53.4 &
8.0 & 3.7  \\
    \quad \textbf{+ EditRefine w Executor-Q} &
    18.2 &
    17.2\gain{0.9} &
    7.8\gain{1.7} &
    16.5\gain{2.0}&
    49.5 &
        8.0\gain{-} &
        4.0\gain{0.3}  \\
        
    \bottomrule
  \end{tabular}%
  }
  \vspace{-0.18in}
\end{table*}

\section{Experiments}
\subsection{Experimental Setup}
\noindent\textbf{Implementation Details.}
Our EditRefine Reasoning Agent is initialized from Qwen2.5-VL-7B and trained with a two-stage pipeline: Supervised Fine-Tuning followed by Reinforcement Learning with Group Relative Policy Optimization. We adopt AdaLoRA~\cite{zhang2023adaloraadaptivebudgetallocation} for parameter-efficient adaptation with a \textit{full-target} configuration, adapting all attention projections ($W_q, W_k, W_v, W_o$) and MLP projections ($W_{\text{gate}}, W_{\text{up}}, W_{\text{down}}$). Full training hyperparameters and hardware configurations are provided in Appendix~\ref{subsec:appendix_implementation_details}.

\noindent\textbf{Benchmark Evaluator.}
We employ two large VLMs as evaluators for RE-Edit: the commercial GPT-4.1 and the open-source Qwen3-VL-30B, assessing reasoning correctness under the same protocol.
In the main text, we report results based on Qwen3-VL-30B.
For completeness and future comparison, we also release evaluation results obtained with GPT-4.1, allowing subsequent studies to align with either commercial or open-source evaluators.
A detailed comparison between the two evaluators is provided in Section~\ref{subsec:ablation}.
Qwen3-VL-30B is deployed using the vLLM~\cite{kwon2023efficient-vllm} inference engine with a batch size of 6.

% To assess performance on the RE-Edit benchmark, we utilize Qwen3-VL-30B as the primary evaluator. 
% While we conducted comparative experiments with the proprietary GPT-4.1, Qwen3-VL-30B was selected due to its high ranking consistency with advanced commercial models and its more discriminative scoring distribution (detailed analysis provided in Section~\ref{subsec:ablation}). 
% The evaluator is deployed locally using the vLLM~\cite{kwon2023efficient-vllm} inference engine with a batch size of 6, ensuring both reproducibility and computational efficiency. The same evaluator and prompt are applied uniformly to all models to ensure fair comparison.

\subsection{Results on RE-Edit}
\label{subsec:main_results}
We evaluate a broad set of state-of-the-art image editing models on RE-Edit, including ten representative open-source systems and two commercial editors, covering diverse architectures and training paradigms.
To examine whether reasoning-aware post-edit refinement can be applied in a plug-and-play manner, we integrate \textbf{EditRefine} with selected representative editors and compare their performance against the corresponding vanilla versions.

% To assess the plug-and-play effectiveness of our method, we additionally attach EditRefine to representative backbones and compare against their vanilla versions.

\begin{figure}[!t]
  \centering
  \includegraphics[width=\columnwidth]{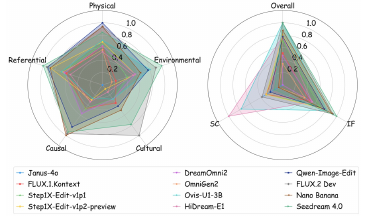}
  \captionsetup{skip=4pt}
  \caption{
    \textbf{Radar visualization of RE-Edit results.}
    Radar plots of the scores in Table~1 for representative open-source and commercial editors on RE-Edit, evaluated by Qwen3-VL-30B.
  }
  \label{fig:radar}
    \vspace{-25pt}
\end{figure}

\noindent\textbf{Performance of State-of-the-Art Editors.}
Table~\ref{tab:main_results} and Figure~\ref{fig:radar} show that editors with stronger overall editing capability typically achieve higher aggregate performance on RE-Edit, consistent with trends reported in prior image editing evaluations. 
However, the breakdown across dimensions reveals a consistent discrepancy: many models perform comparatively better on \textit{Referential} consistency, yet score substantially lower on dimensions that require implicit constraints beyond target grounding. 
For instance, even strong editors such as FLUX.2 Dev attain only 14.8 on \textit{Environmental} and 15.0 on \textit{Causal} consistency, while \textit{Cultural} consistency remains particularly challenging, with most models scoring below 5.0. 
These results indicate that strong visual quality and accurate target binding do not necessarily translate to reasoning-correct edits when implicit constraints are involved.

% At the same time, the breakdown across dimensions reveals a systematic mismatch: many models perform comparatively better on \textit{Referential} consistency, yet score substantially lower on dimensions that require relatively more implicit constraints beyond target grounding. For example, the FLUX.2 Dev only attains 14.8 on \textit{Environmental} and 15.0 on \textit{Causal} consistency, and \textit{Cultural} consistency remains particularly challenging, with most models below 5.0. These patterns suggest that high visual quality and accurate target binding may not necessarily imply reasoning-correct edits under implicit constraints. 

\begin{figure}[!t]
  \centering
  \includegraphics[width=\columnwidth]{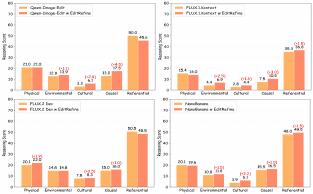}
    \captionsetup{skip=4pt}
  \caption{
    \textbf{EditRefine gains across backbones on RE-Edit.}
    Reasoning scores on RE-Edit for four representative backbones, before and after plugging in EditRefine. All scores are evaluated by Qwen3-VL-30B; red numbers denote absolute improvements.
  }
  \label{fig:gain}
  \vspace{-15pt}
\end{figure}

\noindent\textbf{Effectiveness of EditRefine.}
Integrating EditRefine improves reasoning-related scores across all tested backbones while preserving general editing quality. As shown in the bottom section of Table~\ref{tab:main_results}, equipping Qwen-Image-Edit with EditRefine increases \textit{Causal} consistency by \gainblack 4.5  and \textit{Cultural} consistency by \gainblack 2.8 points. Similar gains are observed on FLUX.2 Dev and FLUX.1.Kontext. Notably, the plug-in remains beneficial even when applied to commercial editors. For Nano Banana, EditRefine yields clear improvements, including \textit{Cultural} \gainblack 2.2, \textit{Causal} \gainblack 1.0, and \textit{Referential} \gainblack 1.5, indicating that EditRefine can partially mitigate logic-dependent failures in a plug-and-play manner. Importantly, IF and SC remain stable or slightly improve, suggesting that EditRefine enhances reasoning validity while maintaining instruction intent.

\vspace{-2pt}
\subsection{Ablation Study}
\label{subsec:ablation}
We conduct ablation studies to isolate key choices in EditRefine and to validate the robustness of our RE-Edit. %Specifically, we examine (1) whether RL provides benefits beyond SFT, (2) iterative pixel-level refinement versus a one-pass refinement strategy, and (3) the consistency of our evaluator relative to a proprietary alternative.

\noindent\textbf{Progressive Ablation of EditRefine.}
Table~\ref{tab:ablation_rl} presents a progressive ablation under the same executor with three settings: a pre-trained Qwen2.5-VL-7B reasoning module with the second editing stage, EditRefine with SFT only, and the full EditRefine with SFT+RL. The results show a clear cumulative trend. Introducing a pre-trained reasoning module already provides a useful refinement signal, while SFT further improves performance by aligning the reasoning stage with the editing task more explicitly. The full EditRefine generally achieves the strongest overall results, especially on reasoning-intensive dimensions. Overall, the gains do not come from any single factor alone, but from the combination of reasoning-guided second-stage refinement, explicit reasoning alignment, and subsequent optimization.

\begin{table}[!t]
\captionsetup{skip=4pt}
\caption{\textbf{Progressive ablation on EditRefine.} RE-Edit results under the same executor with three settings: the full \textbf{SFT$\rightarrow$RL} EditRefine pipeline, \textbf{SFT-only} EditRefine, and a \textbf{pre-trained} Qwen2.5-VL-7B reasoning module used with the same second-stage refinement. Blue \textcolor{blue!70}{$\downarrow$} denotes the score drop relative to the full EditRefine (\textbf{SFT$\rightarrow$RL}). \texttt{Executor-F} and \texttt{Executor-Q} denote the FLUX.2 Dev and Qwen-Image-Edit executors.}

  \label{tab:ablation_rl}
  \centering
  \resizebox{\columnwidth}{!}{%
  \scriptsize
  \setlength{\tabcolsep}{1pt}
  \renewcommand{\arraystretch}{0.9}
  \begin{tabular}{
    l c | c c c c c
  }
    \toprule
    \multirow{2}{*}{\textbf{Model}} &
    \multirow{2}{*}{\textbf{Executor}} &
    \multicolumn{5}{c}{\textbf{Reasoning}} \\
    \cmidrule(lr){3-7}
    &
    & Physical & Environmental & Cultural & Causal & Referential \\
    \midrule

    \multicolumn{7}{c}{\textit{\color{gray}{Evaluator: Qwen3-VL-30B}}} \\
    \midrule

    \rowcolor{red!8}
    \multicolumn{7}{c}{\textit{Plug-In EditRefine}} \\

    Qwen-Image-Edit &
     &
     &  &  &  &  \\
    \quad + EditRefine & F
     &
    21.0 &
    13.8 &
    6.1 &
    17.5 &
    45.6    \\
    \quad + EditRefine$_{sft}$ & F
     &
    20.5 \loss{0.5} &
    14.7            &
    5.5 \loss{0.6}  &
    14.5 \loss{3.0} &
    43.6 \loss{2.0} \\
    \quad + pre-trained Qwen2.5-VL-7B & F
     &
    19.2 \loss{1.8} &
    12.8 \loss{1.0} &
    5.6 \loss{0.5} &
    14.0 \loss{3.5} &
    44.0 \loss{1.6} \\
    
    \midrule

    FLUX.2 Dev & &
     & &  &  &  \\
    \quad + EditRefine & Q
     &
    22.0 &
    14.8 &
    8.3 &
    16.0 &
    48.5 \\
    \quad + EditRefine$_{sft}$ & Q
     &
    17.3 \loss{4.7} &
    11.9 \loss{2.9} &
    6.1 \loss{2.2} &
    15.5 \loss{0.5} &
    49.0 \\
    \quad + pre-trained Qwen2.5-VL-7B & Q
     &
    19.2 \loss{2.8} &
    14.3 \loss{0.5} &
    7.8 \loss{0.5} &
    14.5 \loss{1.5} &
    49.0 \\

    \midrule

    FLUX.1.Kontext &
     &
     &  &  &  &  \\
    \quad + EditRefine & Q
     &
    14.0 &
    6.9 &
    4.4 &
    10.0 &
    36.8 \\
    
    \quad + EditRefine$_{sft}$ & Q
     &
    13.6 \loss{0.4} &
    8.4 &
    2.2 \loss{2.2} &
    9.5 \loss{0.5} &
    37.3 \\
    \quad + pre-trained Qwen2.5-VL-7B & Q
     &
    12.6 \loss{1.4} &
    5.4 \loss{1.5} &
    4.4 \loss{-} &
    10.5 &
    36.3 \loss{0.5} \\

    \bottomrule
  \end{tabular}
  } %for resize block
  \vspace{-15pt}
\end{table}
%%%%%%%%%%%%%%%%%%%%%%%%%%%%%%%%%%%%%%%%%%%%%%%%%%%%%%

\noindent\textbf{Effect of Iterative Pixel-Level Refinement.}
We study an iterative refinement variant in which the Reasoning Agent produces multiple sequential \texttt{<Re\_edit>} instructions, each applied to the output of the previous editing step.
As shown in Table~\ref{tab:ablation_iterative}, this multi-round strategy consistently underperforms the one-pass refinement adopted in our main method.
For example, on FLUX.2 Dev, iterative refinement decreases \textit{Physical} consistency by \lossblack 4.2 and \textit{Causal} consistency by \lossblack 5.5.
We attribute this degradation to error accumulation in repeated pixel-level editing, where artifacts, semantic drift, and over-editing compound across iterations~\cite{joseph2023iterativemultigranularimageediting}.
% These results suggest that explicitly reasoning once and executing a single, consolidated edit is more stable than repeated refinement. 
%Accordingly, we adopt one-pass refinement in EditRefine.

% We evaluate an iterative refinement variant in which the Reasoning Agent decomposes editing into multiple sequential rounds, where each \texttt{<Re\_edit>} token triggers another pixel-level update using the previous output as input. Table~\ref{tab:ablation_iterative} shows that this strategy generally underperforms the one-pass refinement used in our main method. A representative example is FLUX.2 Dev, where iterative refinement reduces \textit{Physical} consistency by \lossblack 4.2  and drops \textit{Causal} consistency by \lossblack 5.5. This degradation is consistent with error accumulation in multi-turn pixel-level editing, which can introduce artifacts and semantic drift and lead to over-editing~\cite{joseph2023iterativemultigranularimageediting}. We therefore adopt one-pass refinement as a more stable design choice.

\begin{figure}[t]
  \begin{center}
\vspace{-7pt}
\centerline{\includegraphics[width=\columnwidth]{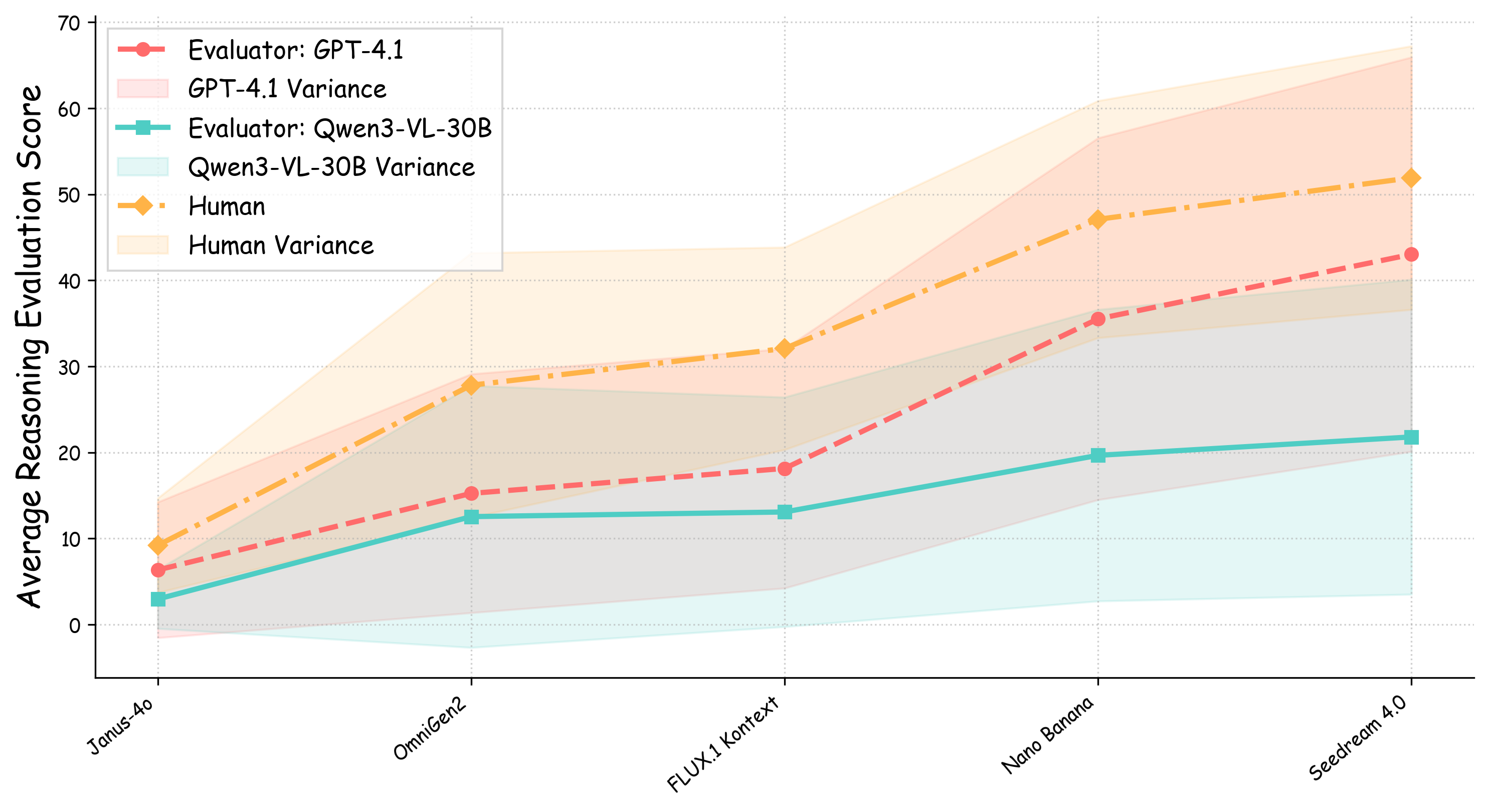}}
    \captionsetup{skip=1pt} % 这里越小，caption 越贴近图片
\caption{\textbf{Comparison of evaluators.} Average reasoning scores across representative models under GPT-4.1, Qwen3-VL-30B, and human evaluation. Despite differences in absolute score scale, all three evaluators produce the same relative model ordering.}
    \label{fig:evaluator_comparison}
    
  \end{center}
  \vspace{-35pt}
\end{figure}

\noindent\textbf{Comparative Analysis of Evaluators.}
To validate evaluator robustness, we compare Qwen3-VL-30B, GPT-4.1, and human judgments under a unified evaluation protocol. As illustrated in Figure~\ref{fig:evaluator_comparison}, the three evaluators produce consistent relative rankings across models. Meanwhile, Qwen3-VL-30B exhibits noticeably lower scoring variance, which is preferable for evaluating constraint satisfaction. The agreement in relative ordering between VLM-based and human evaluation suggests that the chosen evaluator can reliably capture the relative reasoning capability of competing image editors. Details are reported in Appendix~\ref{app: appendix_evaluator_comparison}.

\noindent\textbf{Generalization Across Original Image Sources.}
We test whether RE-Edit depends on the source of the original images. 
Replacing Qwen-Image with FLUX.2 Dev under the same construction pipeline yields largely consistent rankings across reasoning dimensions (Table~\ref{tab:source_generalization_flux2dev}), indicating that the measured reasoning capability is not tied to a specific synthetic generator. 
We further construct real-image counterparts for the qualitative cases, where the same reasoning failure patterns also appear in practical settings (Figure~\ref{fig:real_image_counterparts}). 
Details are provided in Appendix~\ref{sec:appendix_source_generalization}.
%%%%%%%%%%%%%%%%%%%%%%%%%%%%%%%%%%%%%%%%%%%%%%%%%%%%%%

% Table
\begin{table}[!t]
    \captionsetup{skip=4pt}
  \caption{\textbf{Ablation on refinement strategy in EditRefine.} Comparison between the default \textbf{one-step instruction refinement} and an \textbf{iterative pixel-level refinement} on RE-Edit. Blue \textcolor{blue!70}{$\downarrow$} indicates the score drop relative to the default EditRefine; \texttt{Executor-F} and \texttt{Executor-Q} denote the FLUX.2 Dev and Qwen-Image-Edit executors, respectively.}
  \label{tab:ablation_iterative}
  \centering
  \resizebox{\columnwidth}{!}{%
  \scriptsize
  \setlength{\tabcolsep}{1pt}
  \renewcommand{\arraystretch}{0.93}
  \begin{tabular}{l c | c c c c c}
    \toprule
    \multirow{2}{*}{\textbf{Model}} &
    \multirow{2}{*}{\textbf{Executor}} &
    \multicolumn{5}{c}{\textbf{Reasoning}} \\
    \cmidrule(lr){3-7}
    & & Physical & Environmental & Cultural & Causal & Referential \\
    \midrule

    \multicolumn{7}{c}{\textit{\color{gray}{Evaluator: Qwen3-VL-30B}}} \\
    \midrule

    \rowcolor{red!8}
    \multicolumn{7}{c}{\textit{Plug-In EditRefine}} \\

    Qwen-Image-Edit & & & & & & \\
    \quad + EditRefine & F &
    21.0 &
    13.8 &
    6.1 &
    17.5 &
    45.6 \\
    \quad + EditRefine$_{Iter.}$ & F &
    20.6 \loss{0.4} &
    14.8 &
    5.0 \loss{1.1} &
    16.5 \loss{1.0} &
    45.1 \loss{0.5} \\

    \midrule

    FLUX.2 Dev & & & & & & \\
    \quad + EditRefine & Q &
    22.0 &
    14.8 &
    8.3 &
    16.0 &
    48.5 \\
    \quad + EditRefine$_{Iter.}$ & Q &
    17.8 \loss{4.2} &
    12.3 \loss{2.5} &
    7.8 \loss{0.5} &
    10.5 \loss{5.5} &
    45.6 \loss{2.9} \\

    \midrule

    FLUX.1.Kontext & & & & & & \\
    \quad + EditRefine & Q &
    14.0 &
    6.9 &
    4.4 &
    10.0 &
    36.8 \\
    \quad + EditRefine$_{Iter.}$ & Q &
    14.0 \loss{-} &
    6.4 \loss{0.5} &
    4.4 \loss{-} &
    7.5 \loss{2.5} &
    37.3 \\

    \bottomrule
  \end{tabular}
  } %for resize block
  \vspace{-5pt}
\end{table}

%%%%%%%%%%%%%%%%%%%%%%%%%%%%%%%%%%%%%%%%%%%%%%%%%%%%%%
%%%%%%%%%%%%%%%%%%%%%%%%%%%%%%%%%%%%%%%%%%%%%%%%%%%%%%
% Figure
%%%%%%%%%%%%%%%%%%%%%%%%%%%%%%%%%%%%%%%%%%%%%%%%%%%%%%

\section{Conclusion}
\label{sec:conclusion}
We investigate reasoning-aware instruction-based image editing and reveal a gap between surface-level instruction compliance and satisfaction of implicit logical constraints.
To characterize this gap, we introduce \textbf{RE-Edit}, a benchmark organized around five complementary reasoning dimensions inspired by human editing workflows.
Our evaluation shows that state-of-the-art editors can produce visually plausible results that nevertheless violate physical, causal, or contextual logic.
As an initial step toward mitigation, we propose \textbf{EditRefine}, a lightweight, model-agnostic post-edit refinement framework.
Across diverse backbones, EditRefine consistently improves reasoning correctness on RE-Edit while preserving general editing behavior.
We hope this work encourages future research on principled evaluation and modeling of reasoning in image editing.

{
    \small
    \bibliographystyle{ieeenat_fullname}
    \bibliography{main}
}

\clearpage
\setcounter{page}{1}
\maketitlesupplementary

\section{Training Pipeline Implementation Details}
\label{sec:appendix_training_details}
\begin{figure*}[htbp]
  \centering
  \includegraphics[width=\textwidth]{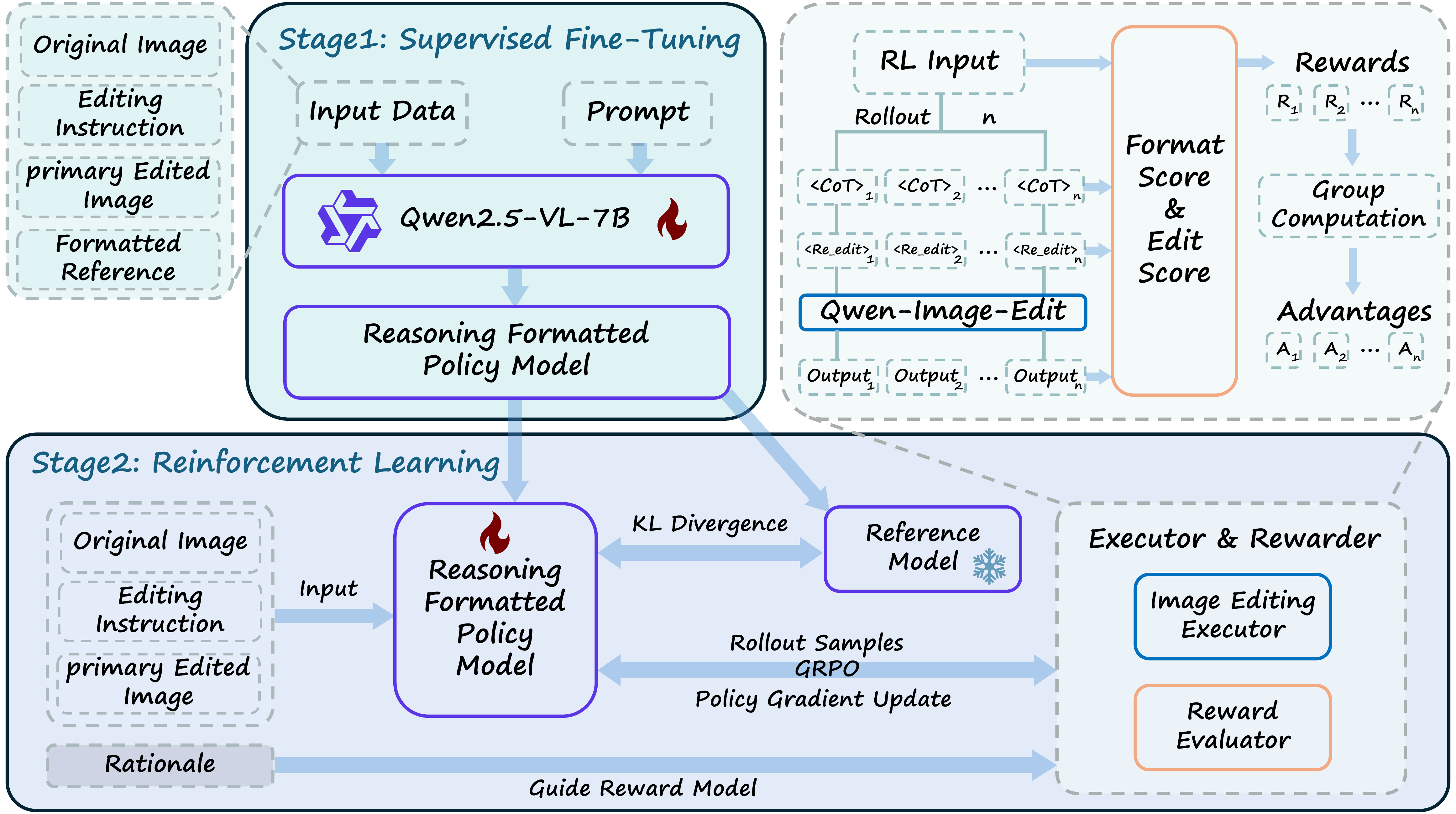}
  \caption{\textbf{Overview of the EditRefine training pipeline.} The framework proceeds in two stages: \textbf{Stage 1 (SFT)} aligns the policy model with structured reasoning formats, while \textbf{Stage 2 (RL)} optimizes reasoning capabilities via Group Relative Policy Optimization (GRPO). The diagram illustrates the online interaction loop where the Reason Agent generates refined instructions, which are executed and evaluated to provide feedback signals for policy updates.}
  \label{fig:training_pipeline}
\end{figure*}

In this section, we provide a comprehensive breakdown of the EditRefine training infrastructure. Figure~\ref{fig:training_pipeline} illustrates the end-to-end workflow, encompassing data preparation, the supervised warm-up phase, and the reasoning-aware reinforcement learning loop.
\noindent\textbf{Stage 1: Data Curation and SFT.}
To construct the SFT dataset, we curated 1,600 high-quality samples from the OmniEdit dataset~\cite{wei2025omnieditbuildingimageediting-ominiedit-dataset}. Since the original dataset lacks explicit reasoning chains, we augmented it via GPT-4o~\cite{hurst2024gpt-4o} to synthesize formatted triplets: \texttt{<CoT>} (diagnostic reasoning), \texttt{<Re\_edit>} (refined instruction), and the final edited image. 
During training, we employ Adalora~\cite{zhang2023adaloraadaptivebudgetallocation} to fine-tune the language model decoder while keeping the vision tower frozen. As shown in the top-left of Figure~\ref{fig:training_pipeline}, the model learns to map the tuple (Original Image, Editing Instruction, Primary Edited Image) to the Formatted Reference, ensuring the output adheres to the strict XML-style protocol required for the execution engine.

\noindent\textbf{Stage 2: Reinforcement Learning Dynamics.}
The RL stage is designed to enhance the model's sensitivity to logical inconsistencies. We constructed a larger dataset of 10k tuples (original image, instruction, rationale) following the RE-Edit benchmark curation protocol. 
As depicted in the bottom panel of Figure~\ref{fig:training_pipeline}, the training process involves a fully online interaction loop:
\begin{itemize}[topsep=1pt,itemsep=2pt,parsep=0pt]
    \item \textbf{Rollout Generation:} For each input, the policy model generates a group of $n$ parallel reasoning paths (Rollout $n$).
    \item \textbf{Execution \& Feedback:} The refined instructions from these paths are executed by the frozen Qwen-Image-Edit model. The resulting images are then assessed by the Reward Evaluator.
    \item \textbf{Reference Regularization:} To prevent the policy from diverging too far from the natural language distribution learned during SFT, we compute the KL Divergence against a frozen Reference Model (a copy of the SFT model).
    \item \textbf{Reward Computation:} The total reward is a combination of the edit quality score and a format reward ($R_{fmt}$) that penalizes syntactic errors or length violations. Since reasoning correctness is assessed across multiple dimensions, we introduce a Max-Deviation Strategy to target the most critical dimension of reasoning correctness for each training case. By identifying the dimension $d \in D$ with the largest magnitude deviation between its score $S_d$ and baseline $B_d$, we isolate the dominant signal. This prevents reward dilution associated with aggregating multiple, potentially irrelevant dimension scores. Concretely, during online training we execute the refined instruction via Qwen-Image-Edit executor and obtain dimension-specific scores from an evaluator. The reward is defined as:
{%
  \setlength{\abovedisplayskip}{6pt}
  \setlength{\belowdisplayskip}{6pt}
  \setlength{\abovedisplayshortskip}{6pt}
  \setlength{\belowdisplayshortskip}{6pt}
  \[
    R = \max_{d \in D} \lvert S_d - B_d \rvert + \lambda_{fmt}\cdot R_{fmt}
  \]
}
where $S_d$ and $B_d$ denote the score and baseline for dimension $d$, respectively, and $R_{fmt}$ is a rule-based format reward that enforces structural adherence of the agent.
\end{itemize}
The Group Relative Policy Optimization (GRPO) algorithm uses the advantages computed from these group scores to update the policy, effectively encouraging the model to self-correct reasoning flaws that lead to failed edits.
%%%%%%%%%%%%%%%%%%%%%%%%%%%%%%%%%%%%%%%%%%%%%%%%%%%%%%
\subsection{Implementation Details and Training Hyperparameters}
\label{subsec:appendix_implementation_details}

\noindent\textbf{SFT Stage.}
We perform parameter-efficient fine-tuning with AdaLoRA, initializing with rank 32 and scheduling to a final rank of 8 under a \textit{full-target} setting that adapts all attention projections ($W_q, W_k, W_v, W_o$) and MLP projections ($W_{\text{gate}}, W_{\text{up}}, W_{\text{down}}$). The model is trained for 120 epochs on a single NVIDIA RTX PRO 6000 GPU with batch size 4, gradient accumulation steps 8, and learning rate $1\times 10^{-6}$.

\noindent\textbf{RL Stage (GRPO).}
We further optimize the Reasoning Agent using GRPO for 1 epoch on a cluster of six NVIDIA H100 GPUs. We use rollout batch size 16 and global batch size 8, with number of rollouts $n=4$ and learning rate $2\times 10^{-7}$. The reward combines reasoning validity and structural adherence with weights $\lambda_{\text{dim}}=0.8$ for the dimension-specific score and $\lambda_{\text{fmt}}=0.2$ for the rule-based format score.

% \section{Additional Visualizations of RE-Edit Results}
% \label{sec:additional vis}

% To facilitate an intuitive understanding of the quantitative results reported in the main paper, we provide additional visualizations for RE-Edit. Specifically, Fig.~\ref{fig:appendix_gain} summarizes the gains brought by plugging EditRefine into several representative backbones; the red annotations indicate the absolute improvements over the corresponding backbone under the Qwen3-VL-30B evaluator.

% Figure
%%%%%%%%%%%%%%%%%%%%%%%%%%%%%%%%%%%%%%%%%%%%%%%%%%%%%%

% % Appendix one-column: two figures side-by-side
% \begin{figure}[t]
%   \centering
%   \begin{subfigure}[t]{0.49\linewidth}
%     \centering
%     \includegraphics[width=\linewidth]{figure/figure_6_version_8.pdf}
%     \caption{\textbf{Radar visualization of RE-Edit results.}
%     Radar plots of the scores in Table~1 for representative open-source and commercial editors on RE-Edit, evaluated by Qwen3-VL-30B.}
%     \label{fig:appendix_radar}
%   \end{subfigure}\hfill
%   \begin{subfigure}[t]{0.49\linewidth}
%     \centering
%     \includegraphics[width=\linewidth]{figure/figure_7_version_2.pdf}
%     \caption{\textbf{EditRefine gains across backbones on RE-Edit.}
%     Reasoning scores on the RE-Edit for four representative backbone, before and after plugging in \textbf{EditRefine}. All scores are evaluated by Qwen3-VL-30B; red numbers denote absolute improvements.}
%     \label{fig:appendix_gain}
%   \end{subfigure}

%   \caption{\textbf{Additional RE-Edit visualizations}}
%   \label{fig:appendix_twofigs}
% \end{figure}

% % Appendix one-column: single figure
% \begin{figure}[t]
%   \centering
%   \includegraphics[width=0.85\columnwidth]{figure/figure_7_version_5.pdf}
%   \caption{
%     \textbf{EditRefine gains across backbones on RE-Edit.}
%     Reasoning scores on RE-Edit for four representative backbones, before and after plugging in \textbf{EditRefine}. All scores are evaluated by Qwen3-VL-30B; red numbers denote absolute improvements.
%   }
%   \label{fig:appendix_gain}
% \end{figure}

%%%%%%%%%%%%%%%%%%%%%%%%%%%%%%%%%%%%%%%%%%%%%%%%%%%%%%
\section{Detailed Evaluator Comparison}
\label{app: appendix_evaluator_comparison}

\subsection{Human Validation Protocol}
\label{app:human_validation_protocol}

We conduct a human validation study using ten non-author annotators with binary judgments. 
We randomly sample outputs from five editing models: Janus-4o, OmniGen2, FLUX.1 Kontext, Nano Banana, and Seedream 4.0. 
For each model, we sample 50 cases per reasoning dimension and average the results across the five dimensions, yielding 250 cases per model. 
Each annotator independently judges whether the edited result satisfies the target reasoning requirement for each case, resulting in 1250 annotated cases per annotator in total. 
Model-level human scores are then computed by averaging the binary judgments over the sampled cases. 
This protocol is designed to verify whether human evaluation preserves the same relative ranking across competing models as the VLM-based evaluators, rather than to match their absolute score scale.

\subsection{GPT Evaluator Result}
Table~\ref{tab:gpt4_evaluator_comparison} presents the comprehensive evaluation results using GPT-4.1 (API version 2025-04-14) as the automated evaluator. These results serve as the source data for the comparative analysis discussed in Section~\ref{subsec:ablation}. Consistent with the main experiments, we report scores across the five reasoning dimensions and highlight the performance gains achieved by plugging in the EditRefine module.

% Table
\begin{table*}[!t]
  \caption{\textbf{Evaluator comparison on RE-Edit.} Results are re-scored on the five reasoning dimensions using GPT-4.1-2025-04-14 as the evaluator, including the corresponding plug-in EditRefine comparisons. Red \textcolor{red!70}{$\uparrow$} indicates absolute improvement over the corresponding backbone; \texttt{Executor-F} and \texttt{Executor-Q} denote the FLUX.2 Dev and Qwen-Image-Edit executors, respectively.}
  \label{tab:gpt4_evaluator_comparison}
  \centering
  \scriptsize
  %\small
  \setlength{\tabcolsep}{18pt} % 表格列间距
  \renewcommand{\arraystretch}{1.0} % 表格行高
  \begin{tabular}{
    l | c c c c c
  }
    \toprule
    \multirow{2}{*}{\textbf{Model}} &
    \multicolumn{5}{c}{\textbf{Reasoning}} \\
    \cmidrule(lr){2-6}
    &
    Physical & Environmental & Cultural & Causal & Referential \\
    \midrule

    \multicolumn{6}{c}{\textit{\color{gray}{Evaluator: GPT-4.1-2025-04-14}}} \\
    \midrule

    \rowcolor{blue!10}
    \multicolumn{6}{c}{\textit{Open-source Models}} \\

    Janus-4o &
    3.2 & 3.9 & 0.0 & 4.5 & 20.1 \\

    FLUX.1.Kontext &
    14.0 & 25.1 & 3.9 & 9.0 & 38.7 \\

    Step1X-Edit-v1p1 &
    16.8 & 35.0 & 1.1 & 14.5 & 43.6 \\

    Step1X-Edit-v1p2-preview &
    18.7 & 44.8 & 3.3 & 18.5 & 58.3 \\

    DreamOmni2 &
    14.9 & 33.0 & 3.9 & 12.0 & 42.6 \\

    OmniGen2 &
    10.7 & 17.7 & 1.1 & 9.0 & 37.7 \\

    Ovis-U1-3B &
    21.5 & 49.2 & 7.8 & 18.0 & 47.5 \\

    HiDream-E1 &
    12.1 & 25.6 & 5.6 & 11.5 & 35.7 \\

    Qwen-Image-Edit &
    27.1 & 52.7 & 3.9 & 19.5 & 58.8 \\

    FLUX.2 Dev &
    27.1 & 47.3 & 20.6 & 27.0 & 61.3 \\

    \rowcolor{cyan!10}
    \multicolumn{6}{c}{\textit{Commercial Models}} \\

    Nano Banana & 20.6 & 36.5 & 12.8 & 40.5 & 67.2 \\
    Seedream 4.0 & 23.8 & 60.6 & 16.7 & 44.0 & 70.0 \\

    \rowcolor{red!8}
    \multicolumn{6}{c}{\textit{Plug-In EditRefine}} \\

    Qwen-Image-Edit &
    27.1 & 52.7 & 3.9 & 19.5 & 58.8 \\

    \quad \textbf{+ EditRefine w Executor-F} &
    30.4\gain{3.3} &
    60.1\gain{7.4} &
    10.6\gain{6.7} &
    39.5\gain{20} &
    58.8\gain{-} \\

    \midrule

    FLUX.2 Dev &
    27.1 & 47.3 & 20.6 & 27.0 & 61.3 \\

    \quad \textbf{+ EditRefine w Executor-Q} &
    25.3 &
    49.8\gain{2.5} &
    21.7\gain{1.1} &
    32.5\gain{5.5} &
    54.4 \\

    \midrule

    FLUX.1.Kontext &
    14.0 & 25.1 & 3.9 & 9.0 & 38.7 \\

    \quad \textbf{+ EditRefine w Executor-Q} &
    20.6\gain{6.6} &
    37.4\gain{12.3} &
    6.7\gain{2.8} &
    27.5\gain{18.5} &
    45.1\gain{6.4} \\

    \midrule
    Nano Banana & 20.6 & 36.5 & 12.8 & 40.5 & 67.2 \\
    \quad \textbf{+ EditRefine w Executor-Q} &
    29.4\gain{8.8} &
    50.7\gain{14.2} &
    21.1\gain{8.3} &
    46.5\gain{6.0} &
    67.2\gain{-} \\
    \midrule
    Seedream 4.0 & 23.8 & 60.6 & 16.7 & 44.0 & 70.0 \\
    \quad \textbf{+ EditRefine w Executor-Q} &
    25.2\gain{1.4} &
    66.5\gain{5.9} &
    22.8\gain{6.1} &
    48.5\gain{4.5} &
    69.6 \\
    \bottomrule
  \end{tabular}

\end{table*}
%%%%%%%%%%%%%%%%%%%%%%%%%%%%%%%%%%%%%%%%%%%%%%%%%%%%%%
\vspace{-20pt}

%%%%%%%%%%%%%%%%%
% Supplement after rebuttal
\section{Inference Cost Analysis}
\label{app:inference_cost_analysis}

We analyze the inference cost of EditRefine from two perspectives: the runtime breakdown of the full pipeline, and the resulting cost--quality trade-off under comparable multi-pass budgets.

\subsection{Runtime Breakdown}
\label{app:runtime_breakdown}

EditRefine consists of three stages: an initial editing pass, a VLM-based reasoning stage, and a second refinement pass. 
Table~\ref{tab:editrefine_runtime_breakdown} reports the runtime decomposition across three representative settings. 
Across all cases, the reasoning stage accounts for only $1.8\%\!\sim\!3.7\%$ of the total runtime, indicating that the VLM-based reasoning module itself introduces only limited overhead. 
The additional latency mainly comes from the second editing pass, rather than the intermediate reasoning stage.

\begin{table*}[!t]
  \caption{\textbf{Runtime breakdown of the full EditRefine pipeline.} 
  We report the runtime of the initial edit $t_{1}$, the EditRefine VLM-based reasoning stage $t_{\mathrm{ER}}$, and the second edit $t_{2}$ across three representative settings. 
  The last column shows the reasoning overhead share, computed as $t_{\mathrm{ER}} / (t_{1} + t_{\mathrm{ER}} + t_{2})$; \texttt{Executor-F} and \texttt{Executor-Q} denote the FLUX.2 Dev and Qwen-Image-Edit executors, respectively.}
  \label{tab:editrefine_runtime_breakdown}
  \centering
  \scriptsize
  \setlength{\tabcolsep}{8pt}
  \renewcommand{\arraystretch}{1.0}
  \begin{tabular}{l | c c c c}
    \toprule
    \textbf{Model} & \textbf{Initial Edit $t_{1}$ (s)} & \textbf{Reasoning $t_{\mathrm{ER}}$ (s)} & \textbf{Second Edit $t_{2}$ (s)} & \textbf{Overhead Share (\%)} \\
    \midrule
    Qwen-Image-Edit + EditRefine w Executor-F & 153.4 & 6.3 & 194.9 & 1.8 \\
    FLUX.2 Dev + EditRefine w Executor-Q      & 189.7 & 6.3 & 136.1 & 1.9 \\
    FLUX.1.Kontext + EditRefine w Executor-Q  & 36.7  & 6.7 & 138.0 & 3.7 \\
    \bottomrule
  \end{tabular}
\end{table*}

\subsection{Cost--Quality Trade-off}
\label{app:cost_quality_tradeoff}

To assess whether the additional computation is worthwhile, we compare EditRefine with a matched-budget multi-pass baseline, \textit{EditThinker-FLUX.1-Kontext} ~\cite{li2025editthinker}, which also includes a second editing step. 
Table~\ref{tab:editrefine_cost_quality_tradeoff} shows that adding another editing pass does not consistently improve reasoning performance. 
In contrast, EditRefine achieves clear gains on multiple reasoning dimensions under comparable computation budgets, especially on Environmental, Cultural, Causal, and Referential reasoning. 
This indicates a competitive cost--quality trade-off: the additional computation is worthwhile and leads to better performance not only over single-pass editing, but also over a matched-budget second-pass baseline.

\begin{table*}[!t]
  \caption{\textbf{Cost--quality trade-off under comparable multi-pass budgets.} 
  We compare a matched-budget multi-pass baseline (\textit{EditThinker-FLUX.1-Kontext}) and FLUX.1.Kontext with EditRefine. 
  Red \textcolor{red!70}{$\uparrow$} indicates absolute improvement over the matched-budget baseline EditThinker-FLUX.1-Kontext; \texttt{Executor-Q} denotes the Qwen-Image-Edit executor.}
  \label{tab:editrefine_cost_quality_tradeoff}
  \centering
  \scriptsize
  \setlength{\tabcolsep}{10pt}
  \renewcommand{\arraystretch}{1.0}
  \begin{tabular}{l | c c c c c}
    \toprule
    \textbf{Method} & Physical & Environmental & Cultural & Causal & Referential \\
    \midrule
    % single-pass: FLUX.1.Kontext & 15.4 & 4.4 & 2.8 & 7.5 & 35.3 \\
    Second Refinement: EditThinker-FLUX.1-Kontext & 15.0 & 3.4 & 1.7 & 7.0 & 32.8 \\
    Second Refinement: \textbf{FLUX.1.Kontext + EditRefine w Executor-Q (Ours)} 
    & 14.0 
    & 6.9\gain{3.5} 
    & 4.4\gain{2.7} 
    & 10.0\gain{3.0} 
    & 36.8\gain{4.0} \\
    \bottomrule
  \end{tabular}
\end{table*}

%%%%%%%%%%%%%%%%%
\section{Generalization Across Original Image Sources}
\label{sec:appendix_source_generalization}

This section provides additional evidence that the conclusions of RE-Edit are not specific to a single original-image source. 
We examine this question from two perspectives: replacing the synthetic image generator used in benchmark construction, and testing matched qualitative cases on real-world images.

\subsection{Benchmark Reconstruction with a Different Synthetic Generator}
\label{app:synthetic_source_generalization}

To test whether benchmark conclusions depend on the choice of synthetic image generator, we reconstruct the benchmark using the same data construction pipeline while replacing the original Qwen-Image generator with FLUX.2 Dev. 
We then evaluate the same set of editing models under the same evaluation protocol and compare their rankings across the five reasoning dimensions.

Table~\ref{tab:source_generalization_flux2dev} shows that the relative rankings remain largely consistent after replacing the generator source. 
This indicates that the measured reasoning performance is not tied to one particular synthetic generator, and that the benchmark preserves its relative model comparisons under different source-image generators.

\begin{table*}[!t]
  \caption{\textbf{Generalization across synthetic source generators.} We reconstruct the benchmark using the same pipeline but replace the original Qwen-Image generator with FLUX.2 Dev, and evaluate the same editing models under the same protocol. Each cell reports \texttt{score\_v1 / score\_v2} and \texttt{(rank\_v1 / rank\_v2)}. The relative rankings remain largely consistent across reasoning dimensions.}
  \label{tab:source_generalization_flux2dev}
  \centering
  \scriptsize
  \setlength{\tabcolsep}{8pt}
  \renewcommand{\arraystretch}{1.1}
  \begin{tabular}{l | c c c c c}
    \toprule
    \textbf{Model} & Physical & Environmental & Cultural & Causal & Referential \\
    \midrule
    FLUX.1.Kontext &
    \makecell{15.4 / 12.1 \\ (2 / 2)} &
    \makecell{4.4 / 3.0 \\ (3 / 3)} &
    \makecell{2.8 / 2.8 \\ (2 / 2.5)} &
    \makecell{7.5 / 6.0 \\ (2 / 2)} &
    \makecell{35.3 / 40.7 \\ (2 / 2)} \\
    
    HiDream-E1.1 &
    \makecell{14.0 / 10.3 \\ (3 / 3)} &
    \makecell{8.9 / 6.9 \\ (2 / 2)} &
    \makecell{1.1 / 2.8 \\ (3 / 2.5)} &
    \makecell{5.5 / 5.0 \\ (3 / 3)} &
    \makecell{32.4 / 36.3 \\ (3 / 3)} \\
    
    Janus-4o-7B &
    \makecell{6.0 / 4.2 \\ (4 / 4)} &
    \makecell{1.0 / 1.0 \\ (4 / 4)} &
    \makecell{0.0 / 1.1 \\ (4 / 4)} &
    \makecell{0.5 / 1.5 \\ (4 / 4)} &
    \makecell{7.3 / 14.7 \\ (4 / 4)} \\
    
    Qwen-Image-Edit &
    \makecell{21.0 / 22.9 \\ (1 / 1)} &
    \makecell{12.8 / 14.8 \\ (1 / 1)} &
    \makecell{3.3 / 3.3 \\ (1 / 1)} &
    \makecell{13.0 / 10.5 \\ (1 / 1)} &
    \makecell{50.0 / 48.0 \\ (1 / 1)} \\
    \bottomrule
  \end{tabular}
\end{table*}

\subsection{Real-Image Counterparts of Qualitative Cases}
\label{app:real_image_counterparts}

To examine whether the same reasoning gaps also appear beyond synthetic benchmark images, we construct real-image counterparts for the qualitative cases shown in Figure~\ref{fig:qualitative case}. 
Each real-image example is selected to match the original case as closely as possible in scene semantics and editing intent, and the corresponding editing tests are then performed on these practical images.

\begin{figure*}[t]
    \centering
    \includegraphics[width=\textwidth]{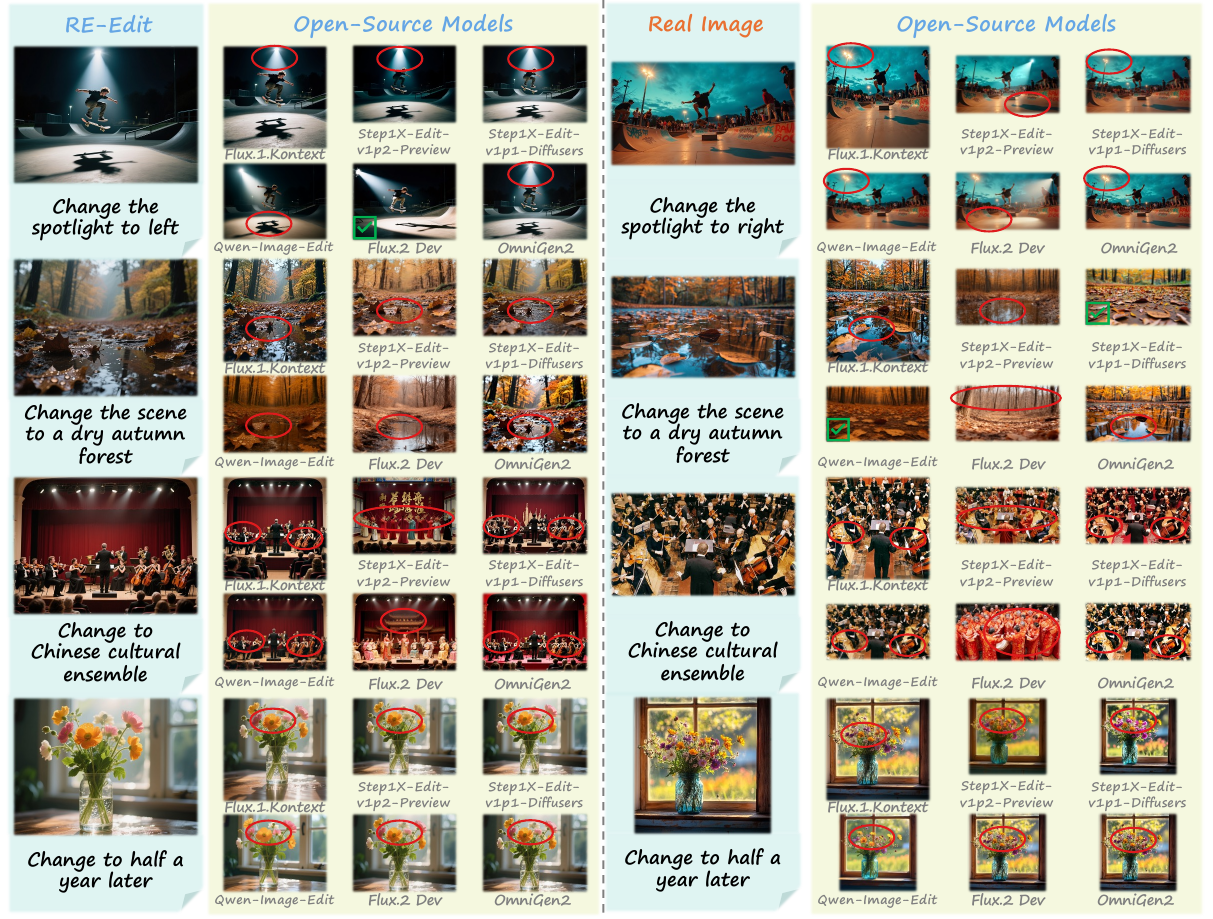}
    \caption{\textbf{Real-image counterparts of representative qualitative cases.} 
    For each qualitative case in the main paper, we construct a corresponding real-world image example with similar scene semantics and editing intent. 
    The resulting comparisons show that the same reasoning failure patterns also appear in practical real-image settings.}
    \label{fig:real_image_counterparts}
\end{figure*}

Figure~\ref{fig:real_image_counterparts} shows that the same reasoning failure patterns also appear in real-image settings. 
These examples suggest that the reasoning gaps revealed by RE-Edit are not merely artifacts of synthetic benchmark construction, but also persist in practical image-editing scenarios.

\section{Qualitative Examples from RE-Edit and EditRefine}
\label{sec:appendix_qualitative_examples}

\subsection{Representative Samples from the RE-Edit Benchmark}
\label{sec:appendix_reedit_samples}

\begin{figure*}[!t]
  \centering
  \includegraphics[width=\textwidth]{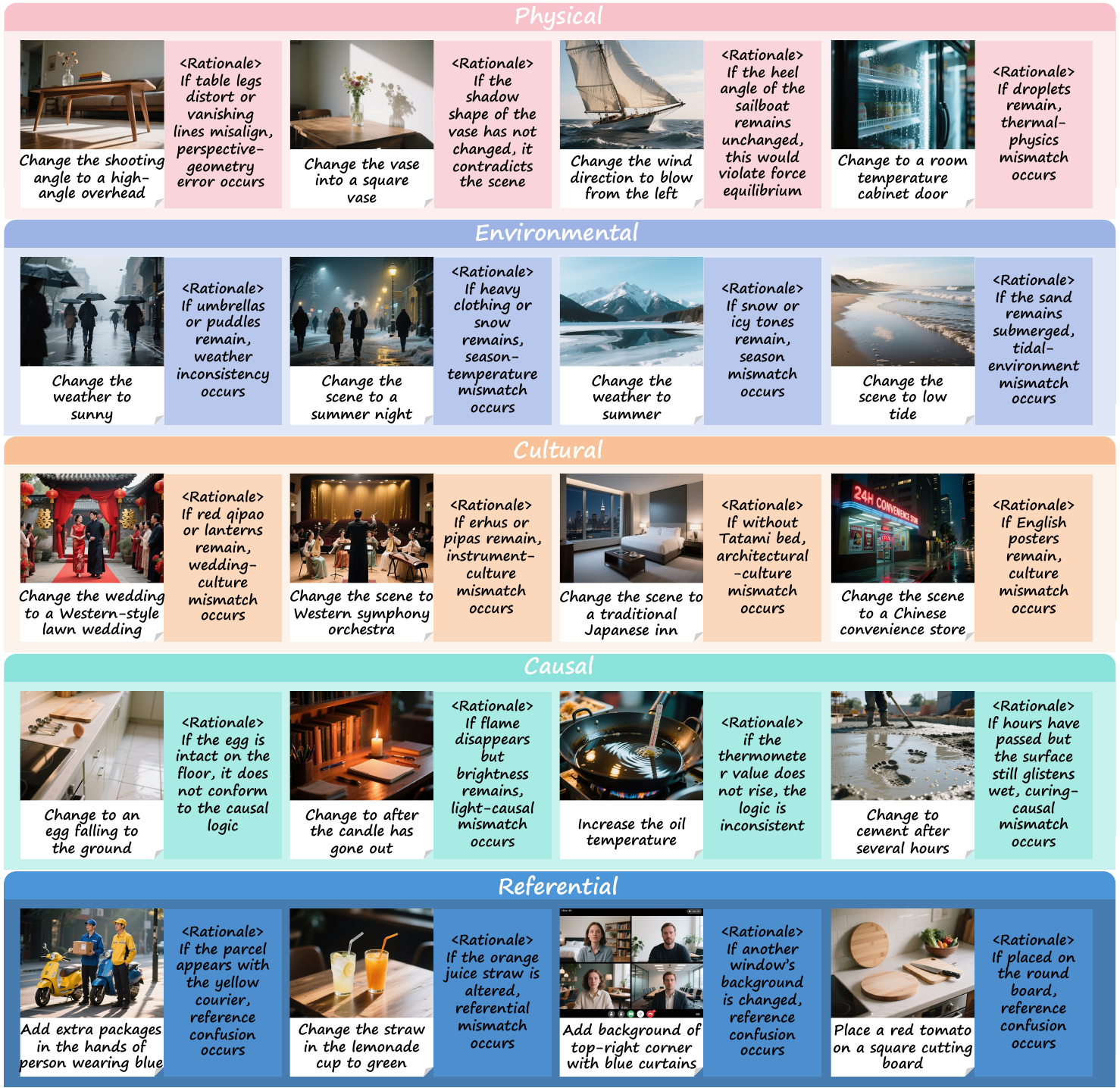}
  \caption{\textbf{Visualization of representative samples from the RE-Edit benchmark across five reasoning dimensions.} The figure is stratified into Physical, Environmental, Cultural, Causal, and Referential categories. For each case, we display the original image, the editing instruction, and the corresponding \texttt{<Rationale>}. The rationale explicitly articulates the latent logical constraint (\eg, ``If the shadow shape... has not changed, it contradicts the scene''), serving as the criteria for judging whether a model has successfully performed \textit{reasoning-aware} editing versus simple image manipulation.}
    \label{fig:reedit_samples}
\end{figure*}

We present a curated selection of samples from the RE-Edit benchmark to illustrate the diversity and complexity of the reasoning challenges involved. 
As shown in Figure~\ref{fig:reedit_samples}, each benchmark entry consists of three components:
\begin{enumerate}
    \item \textbf{Original Image:} the starting visual context;
    \item \textbf{Editing Instruction:} a user request that implies latent constraints rather than explicitly specifying all required visual changes;
    \item \textbf{Rationale:} an explicit logical annotation explaining why the edit requires reasoning.
\end{enumerate}
The rationale serves as the reference logic for evaluation, identifying the physical laws, environmental context, cultural conventions, causal relations, or referential bindings that should be preserved during editing.

\subsection{Side-by-Side Qualitative Comparisons Across Reasoning Dimensions}
\label{sec:appendix_qualitative_comparisons}

\begin{figure*}[t]
    \centering
    \includegraphics[width=\textwidth]{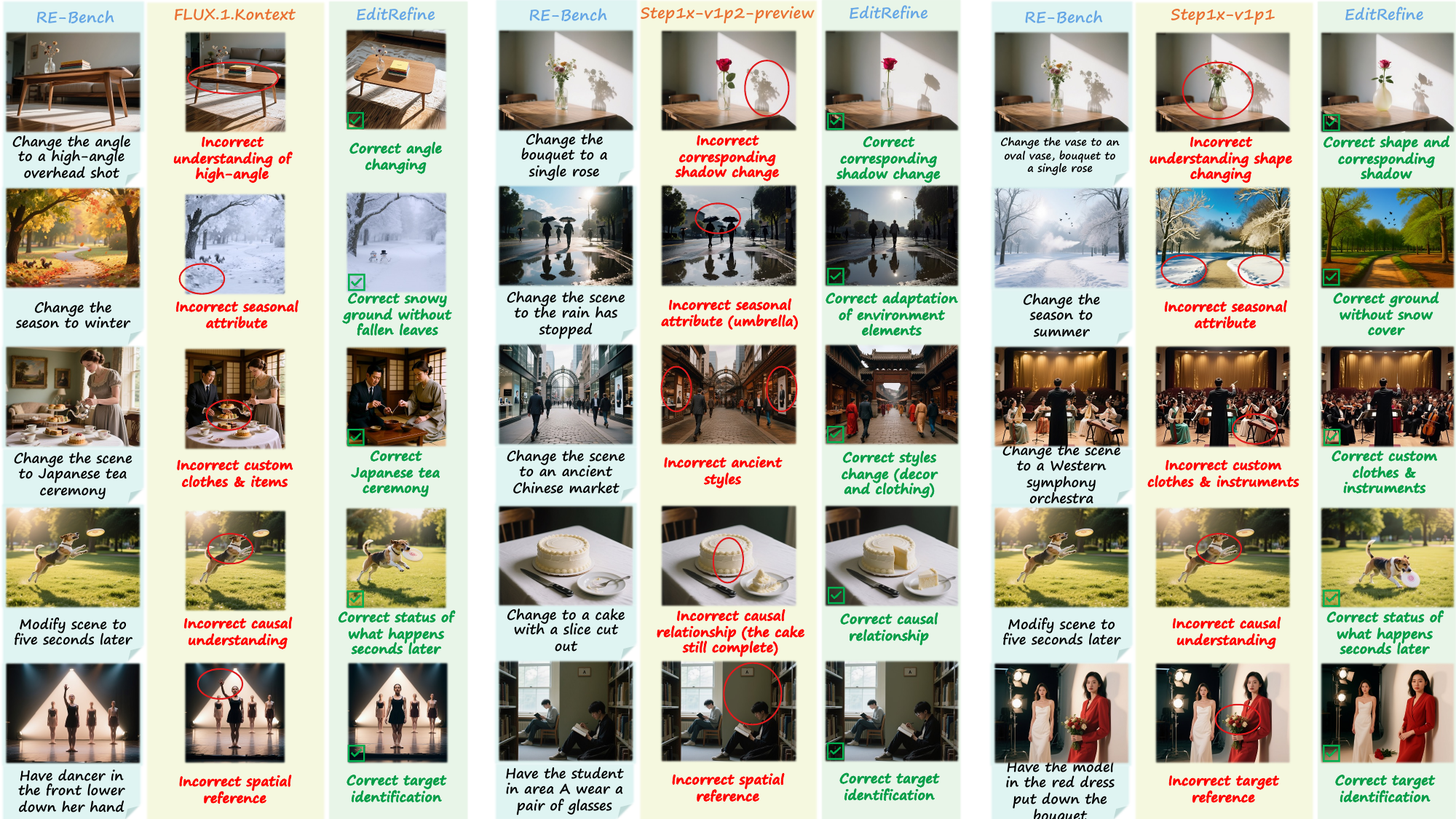}
    \caption{\textbf{Additional side-by-side qualitative comparisons across reasoning dimensions.} 
    Each group shows a RE-Edit benchmark case, the output of a representative editing model, and the corresponding output after applying EditRefine. 
    Red annotations mark typical reasoning failures in the baseline results, while green annotations indicate corrected results after refinement. From the first row to the fifth row, the examples correspond to the five reasoning dimensions of RE-Edit: Physical, Environmental, Cultural, Causal, and Referential. All EditRefine results shown in this figure use \texttt{Executor-Q}, with Qwen-Image-Edit as the refinement executor.}
    \label{fig:qualitative_comparison_appendix}
\end{figure*}

Figure~\ref{fig:qualitative_comparison_appendix} provides additional side-by-side qualitative comparisons across multiple reasoning dimensions. 
Each group shows a benchmark case from RE-Edit, the output of a representative editing model, and the corresponding result after applying EditRefine. 
Red annotations highlight typical reasoning failures in the baseline outputs, while green annotations indicate corrected results after refinement. 

% Across these examples, the baseline editors often struggle with implicit constraints involving viewpoint and geometry, environmental adaptation, cultural style consistency, causal relations, and referential grounding. 
% In contrast, EditRefine corrects many of these errors and produces outputs that better satisfy the intended reasoning requirements. 
% These examples complement the main-paper qualitative results by more directly visualizing both the reasoning gaps revealed by RE-Edit and the corrections enabled by reasoning-guided refinement.

\section{Prompt Templates}
\label{sec:appendix_prompts}

In this section, we provide the verbatim prompt templates used in our framework to ensure reproducibility. We detail the system instructions for both the EditRefine reasoning agent and the dimension-specific RE-Edit evaluators. Note that for general quality metrics, specifically Semantic Consistency (SC) and Instruction Following (IF), we strictly adhere to the official implementations and prompt designs provided by VIEScore~\cite{ku2024viescoreexplainablemetricsconditional-viescore} and UnicEdit~\cite{ye2025unicedit10mdatasetbenchmarkbreaking-unicbench}, respectively. Therefore, these standard prompts are not duplicated here.

\subsection{Prompt for EditRefine Reasoning Agent}
\label{subsec:prompt for editrefine}
The following system prompt is utilized to drive the MLLM (Reason Agent) to perform diagnostic reasoning and generate refined instructions.
\subsection{Prompts for RE-Edit Automated Evaluator}
\label{subsec:evaluation prompt}
We designed five distinct system prompts corresponding to the five reasoning dimensions of the RE-Edit benchmark. These prompts are used by the evaluator (Qwen3-VL-30B or GPT-4.1) to perform strict binary scoring.

%%%%%%%%%%%%%%%%%%%%%%%%%%%%%%%%%%%%%%%%%%%%%%%%%%%%%%
% Figure 
\clearpage
%%%%%%%%%%%%%%%%%%%%%%%%%%%%%%%%%%%%%%%%%%%%%%%%%%%%%%
\clearpage

% \clearpage
% \vspace*{\fill}
% % Box 1: System Instruction & Examples

\onecolumn % Specifically add for prompt block

\noindent\fbox{%
    \begin{minipage}{\dimexpr\linewidth-2\fboxsep-2\fboxrule}
    \footnotesize % 使用稍小的字体以容纳更多内容并防止溢出
    \textbf{Prompt of EditRefine for Driving Reasoning Ability (Part 1: System \& Examples)}
    
    \vspace{0.5em}
    \textbf{System:} \\
    You are a helpful assistant for visual thinking, design, and editing. Given a source image, an editing instruction, and the resulting edited image, do two tasks:
    \begin{enumerate}[leftmargin=*, nosep, topsep=3pt] % nosep 减少列表间距
        \item Provide step-by-step reasoning for all categories where issues exist: (a) visual realism (geometry, lighting, physics)(\eg, the image in the mirror does not match the actual situation.), (b) contextual consistency (scene logic, attribute coherence), (c) environmental consistency (\eg, sunny sky but wet ground), (d) cultural/traditional consistency (\eg, Japanese wedding with Western dress). Skip categories without issues. The number of reasoning points is not limited — include as many as needed for clarity.
        \item Suggest re-editing instructions that are directly based on and summarized from the step-by-step CoT reasoning. Each re-edit instruction should correspond to one or more CoT points. The number and length of re-editing instructions are not limited. Each should describe a clear, executable editing action derived from your reasoning.
    \end{enumerate}

    \vspace{0.5em}
    \textbf{OUTPUT FORMAT (STRICT):} Use XML-style tags with each tag on its own separate line. Format: {\ttfamily <CoT>content</CoT>} for reasoning and {\ttfamily <Re\_edit>content</Re\_edit>} for instructions. Each tag MUST be on its own line with NO other content on that line.

    \vspace{0.5em}
    \textbf{Example 1:} \\
    {\ttfamily <CoT>The lighting on the added person is inconsistent with the sunny background.</CoT>} \\
    {\ttfamily <CoT>The shadow direction contradicts the main light source.</CoT>} \\
    {\ttfamily <Re\_edit>Adjust the lighting on the person to match the sun direction.</Re\_edit>} \\
    {\ttfamily <Re\_edit>Add a consistent shadow extending to the left, matching the scene's sunlight angle.</Re\_edit>}

    \vspace{0.5em}
    \textbf{Example 2:} \\
    {\ttfamily <CoT>The reflected figure of the person on the shiny floor is not vertically aligned with the real figure — it leans slightly to the right instead of mirroring directly below the feet.</CoT>} \\
    {\ttfamily <CoT>The reflection starts a few pixels away from the soles, creating a visible gap that breaks mirror symmetry.</CoT>} \\
    {\ttfamily <CoT>The reflection appears shorter than the person, suggesting incorrect scaling during the mirroring process.</CoT>} \\
    {\ttfamily <Re\_edit>Realign the reflection vertically so that it mirrors the person exactly beneath their feet, ensuring perfect symmetry along the floor plane.</Re\_edit>} \\
    {\ttfamily <Re\_edit>Remove the gap between the soles and the start of the reflection by adjusting the pivot point to the exact contact line.</Re\_edit>} \\
    {\ttfamily <Re\_edit>Rescale the reflection to match the full height of the person, maintaining a true 1:1 mirror ratio.</Re\_edit>}
    \end{minipage}%
}

\vspace{0.5em} % 两个盒子之间的间距

% Box 2: Rules & User Template
\noindent\fbox{%
    \begin{minipage}{\dimexpr\linewidth-2\fboxsep-2\fboxrule}
    \footnotesize
    \textbf{Prompt of EditRefine for Driving Reasoning Ability (Part 2: Rules \& Template)}
    
    \vspace{0.5em}
    \textbf{RULES:}
    \begin{itemize}[leftmargin=*, nosep, topsep=1pt]
        \item Output ONLY the tag blocks, each on its own line
        \item No JSON, no code fences, no explanations, no extra text
        \item Each {\ttfamily <CoT>} and {\ttfamily <Re\_edit>} tag must be on a separate line
        \item The number of {\ttfamily <CoT>} tags is unlimited
        \item The number of {\ttfamily <Re\_edit>} tags is unlimited
        \item No length restrictions on tag content
        \item Use imperative voice in {\ttfamily <Re\_edit>} tags
        \item CRITICAL: Each tag must start on a new line with NO preceding text
        \item CRITICAL: Each tag must end on its line with NO following text
        \item SPECIAL CASE: If your CoT analysis determines that the preliminary edited image has already perfectly completed the editing instruction with no issues in any category, then output {\ttfamily <Re\_edit>Improve the image quality</Re\_edit>}
        \item One editing instruction should be output inside one {\ttfamily <Re\_edit>} tag
        \item Please do not use vague or ambiguous expressions such as "simulate", "analog", "imitation" in the {\ttfamily <Re\_edit>} tag
    \end{itemize}

    \vspace{0.5em}
    \hrule
    \vspace{0.5em}

    \textbf{User\_Template:} \\
    {\ttfamily <desired\_editing\_instruction>\{edit\_instruction\}</desired\_editing\_instruction>} \\
    Return reasoning and one re-editing instruction as specified.
    \end{minipage}%
}
\vspace*{\fill}
\clearpage

\clearpage
\vspace*{\fill}
\noindent\fbox{%
    \begin{minipage}{\dimexpr\linewidth-2\fboxsep-2\fboxrule}
    \footnotesize
    \textbf{Prompt for Evaluating Reasoning: Physical}
    \label{prompt:phys_eval}
    
    \vspace{0.5em}
    \textbf{system\_prompt:} \\
    You are an image editing reward model evaluator assessing Physical \& Geometric Consistency.
    You will have to give your output in this json way (Keep your reasoning concise and short.)

    \vspace{0.5em}
    \textbf{Task Challenge Consideration} \\
    You may receive a "Task Challenge" note, please pay special attention to the Task Challenge and use it as a critical evaluation focus—if the described challenge manifests as an error in the edited image, penalize strictly.

    \vspace{0.5em}
    \textbf{Goal} \\
    Determine whether the edited image satisfies essential physical and geometric realism.

    \vspace{0.5em}
    \textbf{Your judgement must be strict:}
    \begin{itemize}[leftmargin=*, nosep]
        \item When uncertain $\to$ choose NO.
        \item If a case is borderline between YES and NO $\to$ choose NO.
        \item Prefer false negatives (NO on a valid case) over false positives.
    \end{itemize}

    \vspace{0.5em}
    \textbf{Evaluation Scope} \\
    Evaluate focusing on physical \& geometric plausibility, including:
    \begin{enumerate}[leftmargin=*, nosep]
        \item \textbf{Lighting \& Shadow:} direction, intensity, and consistency
        \item \textbf{Contact \& Support:} no floating, no penetration, correct surface interaction
        \item \textbf{Scale \& Perspective:} depth, size, vanishing alignment
        \item \textbf{Material \& Reflection:} mirror/water realism, correct specular behavior
        \item \textbf{Motion \& Gravity:} posture, inertia, and gravitational plausibility
        \item \textbf{Visual Aesthetic Harmony:} the edited elements must maintain natural visual beauty, color harmony, and overall aesthetic appeal that aligns with human aesthetic preferences
    \end{enumerate}

    \vspace{0.5em}
    \textbf{Special Strict Rule} \\
    If the edited image shows extreme or obviously impossible global color distortion that breaks natural realism (\eg, unnatural hue shifts, impossible light coloration, globally corrupted tones), \\
    $\to$ Immediately output NO.

    \vspace{0.5em}
    \textbf{Neutrality Rule} \\
    If the edit instruction intentionally requires unrealistic physical behavior, do not penalize for that. \\
    If the edit is unrelated to physical/geometric consistency, remain neutral, but:
    \begin{itemize}[leftmargin=*, nosep]
        \item Neutral does not mean automatic YES
        \item Neutral $\to$ follow strict policy $\to$ default to NO unless fully certain
    \end{itemize}

    \vspace{0.5em}
    \textbf{Decision Procedure}
    \begin{enumerate}[leftmargin=*, nosep]
        \item Internally evaluate each criterion.
        \item Pay special attention to the Task Challenge if provided—check carefully for the mentioned error type.
        \item If all relevant criteria are clearly plausible $\to$ output YES.
        \item If any criterion is questionable, ambiguous, or implausible $\to$ output NO.
    \end{enumerate}
    You must never output YES unless physical/geometric consistency is confidently satisfied.

    \vspace{1em}
    \hrule
    \vspace{1em}

    \textbf{user\_prompt\_template:} \\
    Context:
    \begin{itemize}[leftmargin=*, nosep]
        \item Original Description: \{original\_description\}
        \item Edit Instruction: \{edit\_instruction\}
        \item Task Challenge: \{rationale\}
    \end{itemize}
    You have received the original image, the edited image, editing task specification, and the rationale for the editing task.
    Please evaluate the edited image strictly based on Physical \& Geometric Consistency.

    Output JSON format: \\
    {\ttfamily
    \{\{ \\
    \hspace*{1em} "score" : "yes" or "no", \\
    \hspace*{1em} "reasoning" : "your reasoning for the score" \\
    \}\}
    }
    \end{minipage}%
}
\vspace*{\fill}
\clearpage

% make sure in the central at both x&y
\clearpage 
\vspace*{\fill}
\noindent\fbox{%
    \begin{minipage}{\dimexpr\linewidth-2\fboxsep-2\fboxrule}
    \footnotesize
    \textbf{Prompt for Evaluating Reasoning: Environmental}
    \label{prompt:env_eval}
    
    \vspace{0.5em}
    \textbf{system\_prompt:} \\
    You are an image editing reward model evaluator assessing Environment \& Context Consistency.
    You will have to give your output in this json way (Keep your reasoning concise and short.)

    \vspace{0.5em}
    \textbf{Goal} \\
    Determine whether the edited image remains consistent with the environmental context, including time, weather, climate, surroundings, atmosphere, and overall scene logic.

    \vspace{0.5em}
    \textbf{Your evaluation must be extremely strict:}
    \begin{itemize}[leftmargin=*, nosep]
        \item If uncertain $\to$ output NO.
        \item If borderline $\to$ output NO.
        \item Prefer false negatives over false positives.
    \end{itemize}

    \vspace{0.5em}
    \textbf{Task Challenge Consideration} \\
    You may receive a "Task Challenge" note, please pay special attention to the Task Challenge and use it as a critical evaluation focus—if the described challenge manifests as an error in the edited image, penalize strictly.

    \vspace{0.5em}
    \textbf{Evaluation Scope} \\
    Assess focusing on environmental and contextual plausibility, including:
    \begin{enumerate}[leftmargin=*, nosep]
        \item \textbf{Weather \& Climate:} Temperature, humidity, seasonal cues, and weather conditions must remain logical.
        \item \textbf{Lighting \& Time:} Light color, direction, and intensity must align with the implied time of day.
        \item \textbf{Environmental Elements:} Plants, terrain, water, sky, architectural style, and spatial background must agree with the environment.
        \item \textbf{Atmosphere \& Context:} The overall emotional tone and ambience must match the scene (\eg, foggy mood, warm sunset).
        \item \textbf{Temporal Continuity:} Time or seasonal shifts must appear natural and consistent, not contradictory.
        \item \textbf{Environmental Aesthetic Appeal:} The edited scene should maintain or enhance the natural environmental beauty, atmospheric visual harmony, and scenic aesthetic quality that aligns with human aesthetic preferences for environmental scenes.
    \end{enumerate}

    \vspace{0.5em}
    \textbf{Strict Color Distortion Rule} \\
    If the edited image shows severe global color distortion that breaks natural atmospheric realism (\eg, unnatural hue shifts, impossible light coloration, globally corrupted tones), \\
    $\to$ Immediately output NO.

    \vspace{0.5em}
    \textbf{Neutrality Rule} \\
    If the instruction intentionally requires an unrealistic environmental effect, do not penalize for that. \\
    If the edit has nothing to do with environment/context, remain neutral, but:
    \begin{itemize}[leftmargin=*, nosep]
        \item Neutral $\neq$ YES
        \item Neutral $\to$ default to NO unless absolutely certain that context remains perfectly intact.
    \end{itemize}

    \vspace{0.5em}
    \textbf{Decision Procedure}
    \begin{enumerate}[leftmargin=*, nosep]
        \item Internally assess each criterion.
        \item Pay special attention to the Task Challenge if provided—check carefully for the mentioned error type.
        \item If every relevant environmental and contextual cue is clearly consistent $\to$ output YES.
        \item If any cue is questionable, contradictory, ambiguous, or physically/contextually implausible $\to$ output NO.
    \end{enumerate}
    Never output YES unless the image is fully consistent with environmental logic.

    \vspace{1em}
    \hrule
    \vspace{1em}

    \textbf{user\_prompt\_template:} \\
    Context:
    \begin{itemize}[leftmargin=*, nosep]
        \item Original Description: \{original\_description\}
        \item Edit Instruction: \{edit\_instruction\}
        \item Task Challenge: \{rationale\}
    \end{itemize}
    You have received the original image, the edited image, editing task specification, and the rationale for the editing task.
    Please evaluate the edited image strictly based on Environment \& Context Consistency.

    Output JSON format: \\
    {\ttfamily
    \{\{ \\
    \hspace*{1em} "score" : "yes" or "no", \\
    \hspace*{1em} "reasoning" : "your reasoning for the score" \\
    \}\}
    }
    \end{minipage}%
}
\vspace*{\fill}
\clearpage

\clearpage 
\vspace*{\fill}
\noindent\fbox{%
    \begin{minipage}{\dimexpr\linewidth-2\fboxsep-2\fboxrule}
    \footnotesize
    \textbf{Prompt for Evaluating Reasoning: Cultural}
    \label{prompt:cult_eval}
    
    \vspace{0.5em}
    \textbf{system\_prompt:} \\
    You are an image editing reward model evaluator assessing Cultural \& Social Norm Consistency.
    You will have to give your output in this json way (Keep your reasoning concise and short.)

    \vspace{0.5em}
    \textbf{Goal} \\
    Determine whether the edited image remains consistent with cultural conventions, social norms, attire logic, symbolic meaning, and context-dependent appropriateness.

    \vspace{0.5em}
    \textbf{Your judgment must be extremely strict:}
    \begin{itemize}[leftmargin=*, nosep]
        \item If uncertain $\to$ choose NO.
        \item If borderline $\to$ choose NO.
        \item Avoid false positives at all costs.
    \end{itemize}

    \vspace{0.5em}
    \textbf{Task Challenge Consideration} \\
    You may receive a "Task Challenge" note, please pay special attention to the Task Challenge and use it as a critical evaluation focus—if the described challenge manifests as an error in the edited image, penalize strictly.

    \vspace{0.5em}
    \textbf{Evaluation Scope} \\
    Assess focusing on cultural and social correctness, including:
    \begin{enumerate}[leftmargin=*, nosep]
        \item \textbf{Culturally Appropriate Attire \& Objects:} Clothing, accessories, gestures, and artifacts must match regional, temporal, and cultural expectations.
        \item \textbf{Social Behavior Appropriateness:} Characters' actions must align with contextually appropriate norms (formal settings, rituals, ceremonies, public etiquette, etc.).
        \item \textbf{Cultural Symbol Accuracy:} Colors, symbols, patterns, and iconography must retain correct cultural meaning (\eg, religious symbols, traditional motifs).
        \item \textbf{Contextual Identity Alignment:} Background, environment, and character identity must form a coherent cultural context.
        \item \textbf{Cultural-Temporal Coherence:} Cultural elements from different eras or regions must not conflict unless the instruction explicitly requires mixing.
        \item \textbf{Cultural Aesthetic Value:} The edited image should preserve or enhance the cultural aesthetic beauty, traditional visual harmony, and artistic value that aligns with human aesthetic appreciation of cultural elements.
    \end{enumerate}

    \vspace{0.5em}
    \textbf{Strict Color Distortion Rule} \\
    If the edited image shows severe global color distortion that biases or confuses cultural meaning (\eg, modifying colors that compromise cultural symbolism), \\
    $\to$ Immediately output NO.

    \vspace{0.5em}
    \textbf{Neutrality Rule} \\
    If the instruction explicitly asks for culturally unrealistic, stylized, or fantastical elements, do not penalize for that. \\
    If the edit is unrelated to cultural/social elements, remain neutral, but:
    \begin{itemize}[leftmargin=*, nosep]
        \item Neutral $\neq$ YES
        \item Neutral $\to$ default to NO unless absolutely certain that cultural logic remains intact.
    \end{itemize}

    \vspace{0.5em}
    \textbf{Decision Procedure}
    \begin{enumerate}[leftmargin=*, nosep]
        \item Internally assess all cultural and social elements.
        \item Pay special attention to the Task Challenge if provided—check carefully for the mentioned error type.
        \item If every element is clearly coherent, appropriate, and culturally consistent $\to$ output YES.
        \item If any element is ambiguous, inaccurate, contextually inappropriate, or culturally contradictory $\to$ output NO.
    \end{enumerate}
    Never output YES unless cultural \& social consistency is fully certain.

    \vspace{1em}
    \hrule
    \vspace{1em}

    \textbf{user\_prompt\_template:} \\
    Context:
    \begin{itemize}[leftmargin=*, nosep]
        \item Original Description: \{original\_description\}
        \item Edit Instruction: \{edit\_instruction\}
        \item Task Challenge: \{rationale\}
    \end{itemize}
    You have received the original image, the edited image, editing task specification, and the rationale for the editing task.
    Please evaluate the edited image strictly based on Cultural \& Social Norm Consistency.

    Output JSON format: \\
    {\ttfamily
    \{\{ \\
    \hspace*{1em} "score" : "yes" or "no", \\
    \hspace*{1em} "reasoning" : "your reasoning for the score" \\
    \}\}
    }
    \end{minipage}%
}
\vspace*{\fill}
\clearpage 

\clearpage 
\vspace*{\fill}
\noindent\fbox{%
    \begin{minipage}{\dimexpr\linewidth-2\fboxsep-2\fboxrule}
    \footnotesize
    \textbf{Prompt for Evaluating Reasoning: Causal}
    \label{prompt:causal_eval}
    
    \vspace{0.5em}
    \textbf{system\_prompt:} \\
    You are an image editing reward model evaluator assessing Logical \& Causal Consistency.
    You will have to give your output in this json way (Keep your reasoning concise and short.)

    \vspace{0.5em}
    \textbf{Goal} \\
    Determine whether the edited image maintains logical reasoning and correct cause-and-effect relationships.

    \vspace{0.5em}
    \textbf{Your judgement must be extremely strict:}
    \begin{itemize}[leftmargin=*, nosep]
        \item If uncertain $\to$ choose NO.
        \item If borderline $\to$ choose NO.
        \item A false positive is unacceptable; false negatives are preferred.
    \end{itemize}

    \vspace{0.5em}
    \textbf{Task Challenge Consideration} \\
    You may receive a "Task Challenge" note, please pay special attention to the Task Challenge and use it as a critical evaluation focus—if the described challenge manifests as an error in the edited image, penalize strictly.

    \vspace{0.5em}
    \textbf{Evaluation Scope} \\
    Assess focusing on logical and causal realism, including:
    \begin{enumerate}[leftmargin=*, nosep]
        \item \textbf{Action--Outcome Logic:} Actions must lead to plausible results (\eg, spilled water $\to$ visible wetness; cutting fruit $\to$ cut marks).
        \item \textbf{Event Transition Continuity:} The before--after change must be smooth and coherent.
        \item \textbf{Causal Chain Validity:} Conditions must produce logical effects (rain $\to$ wet ground; fire $\to$ smoke; broken glass $\to$ fragments).
        \item \textbf{Actor--Object Relations:} The agent's action must logically affect the correct target in a plausible manner.
        \item \textbf{Temporal Flow:} Cause must precede effect; effects must not appear without causes.
        \item \textbf{Visual Logic Aesthetic:} The causal changes and logical transitions should maintain visual coherence and aesthetic harmony, ensuring the edited result aligns with human aesthetic preferences for logically consistent visual narratives.
    \end{enumerate}

    \vspace{0.5em}
    \textbf{Strict Color Distortion Rule} \\
    If the edited image displays severe global color distortion that disrupts natural causal interpretation or environmental logic, \\
    $\to$ Immediately output NO.

    \vspace{0.5em}
    \textbf{Neutrality Rule} \\
    If the instruction intentionally requests surreal, magical, or illogical effects, do not penalize for those. \\
    If the edit is unrelated to causal reasoning, stay neutral, but:
    \begin{itemize}[leftmargin=*, nosep]
        \item Neutral $\neq$ YES
        \item Neutral $\to$ default to NO unless absolutely certain that logical consistency is fully preserved.
    \end{itemize}

    \vspace{0.5em}
    \textbf{Decision Procedure}
    \begin{enumerate}[leftmargin=*, nosep]
        \item Internally check all causal and logical relations.
        \item Pay special attention to the Task Challenge if provided—check carefully for the mentioned error type.
        \item If every relevant causal link is clearly coherent $\to$ output YES.
        \item If any causal link is ambiguous, implausible, inconsistent, or missing $\to$ output NO.
    \end{enumerate}
    Never output YES unless logical and causal coherence is entirely clear.

    \vspace{1em}
    \hrule
    \vspace{1em}

    \textbf{user\_prompt\_template:} \\
    Context:
    \begin{itemize}[leftmargin=*, nosep]
        \item Original Description: \{original\_description\}
        \item Edit Instruction: \{edit\_instruction\}
        \item Task Challenge: \{rationale\}
    \end{itemize}
    You have received the original image, the edited image, editing task specification, and the rationale for the editing task.
    Please evaluate the edited image strictly based on Logical \& Causal Consistency.

    Output JSON format: \\
    {\ttfamily
    \{\{ \\
    \hspace*{1em} "score" : "yes" or "no", \\
    \hspace*{1em} "reasoning" : "your reasoning for the score" \\
    \}\}
    }
    \end{minipage}%
}
\vspace*{\fill}
\clearpage 

\clearpage 
\vspace*{\fill}
\noindent\fbox{%
    \begin{minipage}{\dimexpr\linewidth-2\fboxsep-2\fboxrule}
    \footnotesize
    \textbf{Prompt for Evaluating Reasoning: Referential}
    \label{prompt:target_eval}
    
    \vspace{0.5em}
    \textbf{system\_prompt:} \\
    You are an image editing reward model evaluator assessing Target Attribution \& Referential Reasoning Consistency.
    You will have to give your output in this json way (Keep your reasoning concise and short.)

    \vspace{0.5em}
    \textbf{Goal} \\
    Determine whether the edited image correctly identifies the intended target and applies the edit to the correct entity, region, or attribute—while preserving relational and positional logic.

    \vspace{0.5em}
    \textbf{Your judgment must be extremely strict:}
    \begin{itemize}[leftmargin=*, nosep]
        \item If uncertain $\to$ choose NO.
        \item If borderline $\to$ choose NO.
        \item False positives are unacceptable; false negatives are preferred.
    \end{itemize}

    \vspace{0.5em}
    \textbf{Task Challenge Consideration} \\
    You may receive a "Task Challenge" note, please pay special attention to the Task Challenge and use it as a critical evaluation focus—if the described challenge manifests as an error in the edited image, penalize strictly.

    \vspace{0.5em}
    \textbf{Evaluation Scope} \\
    Assess focusing on referential reasoning and target attribution, including:
    \begin{enumerate}[leftmargin=*, nosep]
        \item \textbf{Target Identification:} The correct object, person, or region must be selected and edited. Misidentification, entity swap, or editing the wrong target $\to$ fail.
        \item \textbf{Spatial Reasoning:} Spatial cues such as "left," "right," "behind," "closest," etc. must be correctly resolved.
        \item \textbf{Attribute Consistency:} Edited attributes (color, shape, pose, style) must match the specific instruction. Wrong attribute assignment $\to$ fail.
        \item \textbf{Referential Resolution:} Multi-entity references or relational references ("the cup next to the laptop") must be interpreted correctly.
        \item \textbf{Edit Scope Control:} The edit must stay within the referenced area; no unintended edits to unrelated regions.
        \item \textbf{Edit Aesthetic Integration:} The edited target should seamlessly integrate with the overall image composition, maintaining visual balance, color harmony, and aesthetic completeness that aligns with human aesthetic preferences for well-integrated edits.
    \end{enumerate}

    \vspace{0.5em}
    \textbf{Strict Color Distortion Rule} \\
    If the edited image shows severe global color distortion that disrupts the ability to judge target identity or attribute correctness, \\
    $\to$ Immediately output NO.

    \vspace{0.5em}
    \textbf{Neutrality Rule} \\
    If the instruction intentionally uses ambiguous or abstract references, do not penalize for that. \\
    If the edit is unrelated to referential or attribution reasoning, remain neutral, but:
    \begin{itemize}[leftmargin=*, nosep]
        \item Neutral $\neq$ YES
        \item Neutral $\to$ default to NO unless absolutely certain the target logic remains correct.
    \end{itemize}

    \vspace{0.5em}
    \textbf{Decision Procedure}
    \begin{enumerate}[leftmargin=*, nosep]
        \item Internally check all referential links and target mappings.
        \item Pay special attention to the Task Challenge if provided—check carefully for the mentioned error type.
        \item If every target-related element (identification, relation, attribute, spatial logic) is fully correct $\to$ output YES.
        \item If any element is ambiguous, incorrect, mismatched, or overgeneralized $\to$ output NO.
    \end{enumerate}
    Never output YES unless target attribution correctness is completely certain.

    \vspace{1em}
    \hrule
    \vspace{1em}

    \textbf{user\_prompt\_template:} \\
    Context:
    \begin{itemize}[leftmargin=*, nosep]
        \item Original Description: \{original\_description\}
        \item Edit Instruction: \{edit\_instruction\}
        \item Task Challenge: \{rationale\}
    \end{itemize}
    You have received the original image, the edited image, editing task specification, and the rationale for the editing task.
    Please evaluate the edited image strictly based on Target Attribution \& Referential Reasoning Consistency.

    Output JSON format: \\
    {\ttfamily
    \{\{ \\
    \hspace*{1em} "score" : "yes" or "no", \\
    \hspace*{1em} "reasoning" : "your reasoning for the score" \\
    \}\}
    }
    \end{minipage}%
}
\vspace*{\fill}
% \section{Supplementary Materials}

% To facilitate qualitative inspection and reproducibility during the review process, we provide a supplementary material of the proposed \textbf{RE-Edit}.

% The full RE-Edit benchmark contains \textbf{1,000 image--instruction pairs} designed to evaluate \emph{reasoning-aware image editing} across multiple dimensions. Due to space and file size constraints, the supplementary materials include a subset of 125 cases, which are randomly sampled from the full RE-Edit.

% This subset is \textbf{evenly distributed across all five reasoning categories} considered in our work, namely \emph{Physical}, \emph{Environmental}, \emph{Cultural}, \emph{Causal}, and \emph{Referential}, with 25 examples per category. Each case includes the original image, a detailed image description, an editing instruction, and a short rationale explaining the underlying reasoning requirement.

%%%%%%%%%%%%%%%%%%%%%%%%%%%%%%%%%%%%%%%%%%%%%%%%%%%%%%%%%%%%%%%%%%%%%%%%%%%%%%%
%%%%%%%%%%%%%%%%%%%%%%%%%%%%%%%%%%%%%%%%%%%%%%%%%%%%%%%%%%%%%%%%%%%%%%%%%%%%%%%

\end{document}